\documentclass[aps,prl,twocolumn,superscriptaddress,amsmath,showpacs,floatfix]{revtex4-1}

\usepackage{graphicx} 
\usepackage{color}

\definecolor{darkgreen}{rgb}{0,0.4,0}

\usepackage[normalem]{ulem} 

\begin{document}


\title{AC Wien effect in spin ice, manifest in non-linear non-equilibrium susceptibility}


\author{V. Kaiser}
\affiliation{Laboratoire de Physique, \'Ecole Normale Sup\'erieure de Lyon, CNRS, 69364 Lyon CEDEX 07, France}
\affiliation{Max-Planck-Institut f\"ur Physik komplexer Systeme, 01187 Dresden, Germany}

\author{S. T. Bramwell}
\affiliation{London Centre for Nanotechnology and Department of Physics and Astronomy, University College London,
London WC1H 0AH, United Kingdom}

\author{P. C. W. Holdsworth}
\affiliation{Laboratoire de Physique, \'Ecole Normale Sup\'erieure de Lyon, CNRS, 69364 Lyon CEDEX 07, France}

\author{R. Moessner}
\affiliation{Max-Planck-Institut f\"ur Physik komplexer Systeme, 01187 Dresden, Germany}


\date{\today}

\begin{abstract}
We predict the non-linear non-equilibrium response of a ``magnetolyte'', the Coulomb fluid of magnetic monopoles in spin ice. This involves an increase of the monopole density due to the second Wien effect---a universal and robust enhancement for Coulomb systems in an external field---which in turn speeds up the magnetization dynamics, manifest in a non-linear susceptibility. Along the way, we gain new insights into the AC version of the classic Wien effect. One striking discovery is that of a frequency window where the Wien effect for magnetolyte and electrolyte are indistinguishable, with the former exhibiting perfect symmetry between the charges. In addition, we find a new low-frequency regime where the growing magnetization counteracts the Wien effect. We discuss for what parameters best to observe the AC Wien effect in Dy$_2$Ti$_2$O$_7$. 
\end{abstract}


\maketitle



\emph{Introduction.---}
Frustrated systems exhibit extensively (quasi-)degenerate ground states~\cite{villain1979,lacroix2011}. This degeneracy facilitates fluctuations, destabilising conventional order and promoting new types of topological states including their exotic quasiparticle excitations.

A striking example is spin ice~\cite{harris1997,bramwell2001,castelnovo2012} in which the low-temperature properties are well described by a Coulomb fluid of magnetic monopoles~\cite{castelnovo2008} (``magnetolyte''). Within the spin description the extensive manifold of low energy states is difficult to treat theoretically. However, the monopole picture provides a framework in which electrolyte theory can be applied and further developed leading to an elegant theoretical description of this exotic frustrated magnet~\cite{castelnovo2008,morris2009,jaubert2011-a,castelnovo2011}. Here we show how the monopole model even gives access to a comprehensive description of the non-linear non-equilibrium response to an external field through the second Wien effect~\cite{onsager1934}.

As one moves beyond linear response, perturbation-driven changes to internal correlations become visible. Non-linear susceptibilities, whether related to higher harmonics or to high-field response, are indispensable macroscopic signatures of this evolution. Notable examples in which non-linear effects are important include optics~\cite{boyd2003}, glassy systems~\cite{fujiki1981,ogielski1985,mezard1991,cugliandolo1997}, liquids~\cite{berthier2011}, superconductors~\cite{rosenstein2010}, heavy fermions~\cite{chandra2013}, and magnets~\cite{wu1993, gingras1997}.

\begin{figure}[htb!]
\centering
\includegraphics{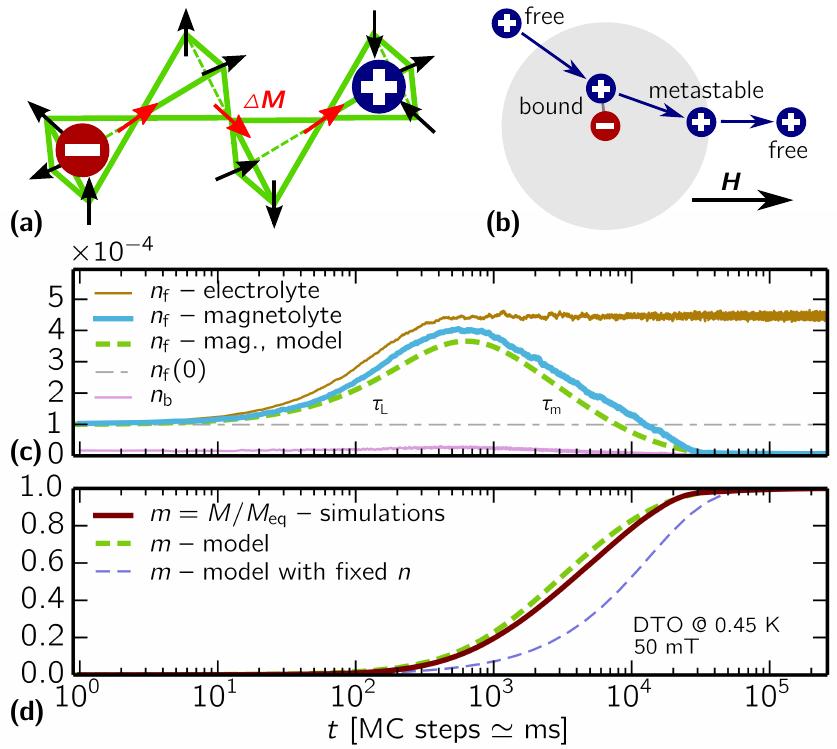}
\caption{\label{fig:quench} (Color online) (a) Monopoles move via spin flips; their current magnetizes the ice manifold; (b) The second Wien effect involves the field enhanced dissociation of bound pairs. Non-linear response: (c) After a field quench, the Wien effect increases the free charge density, $n_\mathrm{f}$, in an electrolyte. In a magnetolyte with the same initial density and temperature, the free monopole density increase is only transient, counteracted by the growing magnetization $m$ of the system (d). The increased monopole density is observable in the faster rate of magnetization $m$ compared to a magnetization process at fixed density $n$. The response is well described by our kinetic model. The bound charge density is only weakly influenced ($n_\mathrm{b}$). Magnetolyte parameters are $T=0.45~\mathrm{K}$, $n_\mathrm{tot}(0)\simeq 1.1\times10^{-4}$, $n_\mathrm{f}(0)\simeq 1.0\times10^{-4}$, and $\mu_0H_0=50~\mathrm{mT}$; electrolyte parameters are set to obtain the same zero field density.}
\end{figure}

Here we show that we can extract the non-linear response of spin ice from the magnetolyte picture over a broad range of temperatures, frequencies and field amplitudes. We use Onsager's exact solution of the second Wien effect~\cite{onsager1934} to predict the full non-linear magnetization response. The underlying mechanism is the generation by the external field of an excess---often very sizeable---of free magnetic charge, which in turn amplifies the magnetic response (Fig.~\ref{fig:quench}a--b).
After introducing the model magnetolyte, we demonstrate that it exhibits the key features of the Wien effect and show how monopole and magnetic responses are coupled. We then develop a kinetic model which accurately reproduces both the Wien effect and the magnetic response observed in simulations.
Finally, we propose an experimental protocol for the non-linear susceptibility and discuss the optimal experimental setting for detecting the AC Wien effect in Dy$_2$Ti$_2$O$_7$.

\emph{Model.---}
Spin ice consists of a network of corner-sharing tetrahedra of Ising-like magnetic moments (spins) constrained to point along the axis connecting the centers of neighboring tetrahedra, which in turn define a diamond lattice of constant $a$. 
Both exchange and dipolar interactions are important. However, for configurations satisfying the ``ice-rules'' of two spins pointing in and two out of each tetrahedron, dipolar interactions maintain an approximate degeneracy which is exact in the dumbbell model, in which spins are replaced by magnetic charge dumbbells that touch at the centers of the tetrahedra. Monopoles on top of this vacuum represent a violation of the ice rules with 3 spins in and 1 out, or 3-out-1-in, corresponding to magnetic charges $Q_\mathrm{m}=\pm 2\mu/a$ (Fig.~\ref{fig:quench}a).
Doubly charged monopoles (4-in or 4-out tetrahedra) are costly in energy and can be neglected over the temperature range considered here \cite{jaubert2011-a}.

The dumbbell model, which we use for both analytics and numerics, leads to a great simplification of the treatment of the dense network of magnetic moments. All the energetics are accounted for by the magnetic Coulomb interaction $U(r) = \pm\mu_0 Q_\mathrm{m}^2 / 4\pi r$~\cite{castelnovo2008} between monopoles along with their chemical potential $\nu$~\cite{jaubert2011-a}. The coupling of the monopoles to the spin background imposes constraints on their motion at short wavelengths, leading e.g.~to a renormalisation of the diffusion constant $D$~\cite{castelnovo2011} by a factor of $\frac{2}{3}$ compared to an unconstrained lattice electrolyte \cite{kaiser2014}. It also accounts for the entropics of spin ice~\cite{ryzhkin2005,moessner2010} as the monopole motion changes the local magnetization. The dumbbell model thus formulates spin ice as a true Coulomb liquid of magnetic monopoles, with the added richness of configurational entropy in the spin background. It has been shown to describe the equilibrium properties of spin ice materials such as $\mathrm{Dy_2Ti_2O_7}$ (DTO) and $\mathrm{Ho_2Ti_2O_7}$ (HTO) in previous studies~\cite{morris2009,fennell2009,jaubert2009,castelnovo2011}.

We simulate the dumbbell model with periodic boundary conditions and Ewald-summed Coulomb interactions~\cite{deleeuw1980}. After equilibration in zero field with both single-spin flip and worm Monte Carlo algorithms~\cite{barkema1998,melko2004,jaubert2011}, the sample evolves via local moves following a chosen field protocol. The locality of the moves ensures that the algorithm maps to physical diffusive dynamics~\cite{kikuchi1991,cheng2006,jaubert2009,sanz2010}. The parameters used in our simulations, and quoted throughout as dimensionful quantities, are extracted from experiments on DTO~\cite{jaubert2009}. 

\begin{figure}[htb!]
\includegraphics{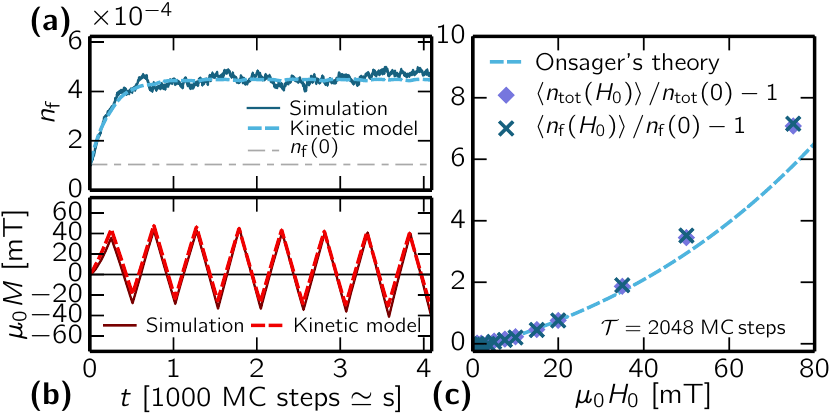}
\caption{\label{fig:square} (Color online) Square wave driving stabilizes the free monopole density increase due to the second Wien effect (a) if the magnetization $M$ stays well below $M_\mathrm{eq}=\chi_T H_0$ (b). The kinetic model captures the response including the transition from equilibrium to a periodic steady state. The amplitude-dependence of the average density increase (c) matches the DC Wien effect theory with no free parameters, confirming spin ice's dynamical ``window'' of electrolyte behavior. Magnetolyte parameters are $T=0.45~\mathrm{K}$, $n_\mathrm{tot}(0)\simeq 1.1\times10^{-4}$, $n_\mathrm{f}(0)\simeq 1.0\times10^{-4}$, and $\mu_0H_0=50~\mathrm{mT}$ (a--b).}
\end{figure}

\begin{figure*}[htb!]
\includegraphics[width=\textwidth]{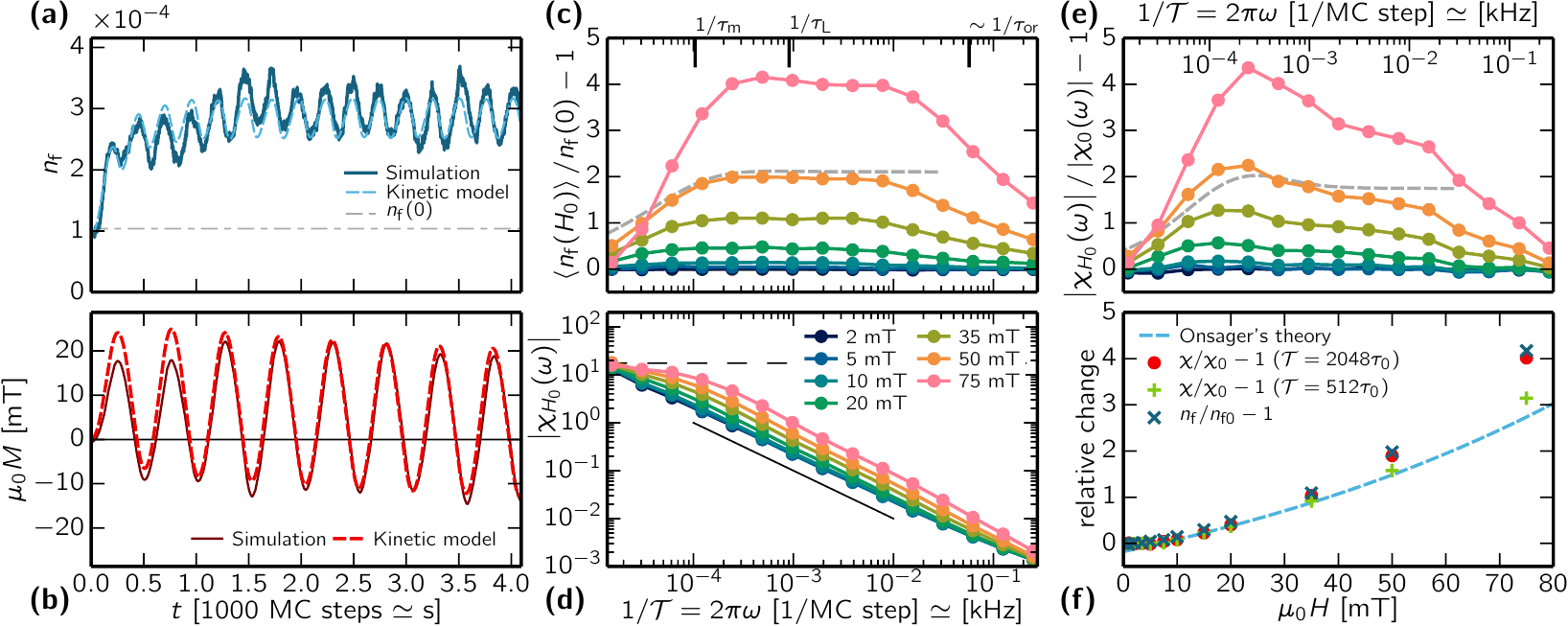}
\caption{\label{fig:sine} (Color online) The free monopole density increase (a) due to sine driving enhances the magnetic response (b). The Wien effect persists over a range of frequencies (c). The enhanced density leads to an increase in the absolute value of the non-linear susceptibility (d); the dashed line is $\chi_T$ and full black line is $\propto1/\omega$. The relative change in $\chi_{H_0}$ is shown in (e) revealing additional features in the Wien effect plateau compared to the density increase. The amplitude dependence stays close to Onsager's theory of the DC Wien effect despite the approximations made (f). The kinetic model (results for $\mu_0H_0=50~\mathrm{mT}$) captures the time evolution of density and magnetization (dashed lines in a--b); the low-frequency transition in density and susceptibility (dashed lines in c and e); and the structure of the susceptibility increase. However, it does not include the high-frequency cutoff due to pair reorientation. Magnetolyte parameters are $T=0.45~\mathrm{K}$, $n_\mathrm{tot}(0)\simeq 1.1\times10^{-4}$, $n_\mathrm{f}(0)\simeq 1.0\times10^{-4}$.}
\end{figure*}

\emph{Wien effect and non-linear response.---}
In weak electrolytes, applying an external electric field strongly increases the density of mobile ions following the enhanced dissociation of bound pairs---the second Wien effect. By analogy one would expect the Wien effect to occur in a weak magnetolyte~\cite{bramwell2009} such as DTO below $\sim1.5~\mathrm{K}$, where $\mu_0 Q_\mathrm{m}^2 / 4\pi a > 2k_\mathrm{B} T$~\cite{bjerrum1926, kobelev2002}. Our first and central result is that our simulations do indeed show the Wien effect in the increase in monopole density (Fig.~\ref{fig:quench}c)! However, this increase is only transient because the monopole currents magnetize the system (Fig.~\ref{fig:quench}d), which eventually halts the Wien effect and even reduces monopole density for unrelated energetic reasons (see Suppl.~Mat.). Crucially, for the temperatures of interest where $n_\mathrm{f}$ is small, the magnetization relaxation time, $\tau_\mathrm{m}$, is longer than the Langevin time lag~\cite{onsager1934}, $\tau_\mathrm{L}$, over which the Wien effect drives the density increase. Hence, periodic switching of the field direction stabilizes the density increase (Fig.~\ref{fig:square}a) when the switching period falls between these two scales, and even for higher frequencies as we discuss in Fig.~\ref{fig:sine}c. The amplitude dependence is staggeringly similar to that for the electrolyte in constant field (Fig.~\ref{fig:square}c).

\emph{Chemical kinetics of monopoles.---}
The backbone of our dynamical analysis is a pair of coupled rate equations, for the monopole density (\ref{eq:chem-lin2}) and for the magnetization (\ref{eq:mag-lin2}) whose derivation we sketch in turn. For the monopole density, it is convenient to start from the field dependent steady state value for a weak electrolyte.

As in the case of weak electrolytes, monopoles in spin ice at low temperatures can be separated into free monopoles and bound (Bjerrum) pairs, treated as distinct chemical species~\cite{bjerrum1926}. Combining this with the creation of bound pairs from the ice manifold (quasi-particle vacuum), we have a double equilibrium: $\textit{vacuum}\rightleftharpoons\textit{bound}\;(n_\mathrm{b})\overset{K}{\rightleftharpoons}\textit{free}\;(n_\mathrm{f}=n_++n_-)$~\cite{onsager1934,persoons1979}. The dissociation constant $K=2\gamma^2n_+n_-/n_\mathrm{b}$ controls the equilibrium between the bound and free charges. The activity coefficient $\gamma$ gives the modification of the free charge density due to correlations, at the mean field level~\cite{debye1923,moore1999}. 

In the second Wien effect, the external field strongly shifts this equilibrium towards the creation of additional free monopoles. The field reorients and dissociates the bound pairs while the bound charge density is swiftly replenished from the vacuum. Onsager~\cite{onsager1934} presented an exact solution for the increase in dissociation rate $k_\mathrm{D}(b)=F(b)k_\mathrm{D}(0)$ and dissociation constant $K(b)=k_\mathrm{D}(b)/k_\mathrm{A}=F(b)K(0)$, where $b=\mu_0^2 Q_\mathrm{m}^3 H_0/8\pi(k_\mathrm{b} T)^2$ is linear in a constant applied field $H_0$, $F(b)=I_1(2\sqrt{2b})/\sqrt{2b} \simeq 1+b+\mathrm{O}(b^2)$ with $I_1$ the modified Bessel function. The association rate stays constant at $k_\mathrm{A}=\chi_T/\tau_0$ \cite{langevin1903}, where $\chi_T=\sqrt{3}\mu_0Q_\mathrm{m}^2/8 k_\mathrm{B}T a$ is the isothermal susceptibility~\cite{ryzhkin2005,jaubert2013}. 

In dilute electrolytes ($n_\mathrm{f}\ll1$), the steady-state free charge density increases as
\begin{equation}\label{eq:onsager}
\Delta n_\mathrm{f}(b)/n_\mathrm{f}(0) = (\gamma(0)/\gamma(b)) \sqrt{F(b)} - 1 \,.
\end{equation}
At fields sufficiently strong to remove the screening atmosphere~\cite{onsager1957,patterson1961}, $\gamma(b) \rightarrow 1$ (``Onsager's theory'', valid above a field of $\sim3\mathrm{mT}$ in Figs.~2c,3f); in lower fields, a crossover in $\gamma(b)$ from unity to the zero-field value occurs. For details of screening, see Ref.~\cite{kaiser2013} and Suppl.~Mat. 

Magnetic monopoles in spin ice react to a force $F=\mu_0 Q_\mathrm{m}(H - M/\chi_T)$~\cite{ryzhkin2005}, with $H$ the internal field along the $[001]$-axis. The term $M/\chi_T$
expresses the entropic bias towards states with low magnetization. Defining $H(t) - M/\chi_T=H_0(h(t)-m)$ with $h(t)=H(t)/H_0$ and $m=M/M_\mathrm{eq}=M/(\chi_T H_0)$, the kinetics of the Wien effect are
\begin{equation}\label{eq:chem-full}
\mathrm{d} n_\mathrm{f}/\mathrm{d} t = k_\mathrm{D}\left(b\left|h(t)-m\right|\right)n_\mathrm{b} - k_\mathrm{A}n_\mathrm{f}^2/2 \,.
\end{equation}
We integrate $n_\mathrm{b}$ out on account of its swift equilibration with the vacuum, whereupon it enters as $(1-n_\mathrm{f})$, representing both bound charges and the monopole vacuum. Further, linearising in $b$ and $n_\mathrm{f}=n_\mathrm{f}^0+\Delta n_\mathrm{f}$ one finds
\begin{equation}\label{eq:chem-lin1}
\mathrm{d} \Delta n_\mathrm{f}/\mathrm{d} t = k_\mathrm{D}(0) b\left|h(t)-m\right| - k_\mathrm{A} n_\mathrm{f}^0\Delta n_\mathrm{f}\;.
\end{equation}
In terms of $\zeta(t)=\Delta n_\mathrm{f}(t) / (b n_\mathrm{f}^0 / 2) $, we obtain our first constitutive equation
\begin{equation}\label{eq:chem-lin2}
\mathrm{d}\zeta/\mathrm{d}t = \left( \left|h(t)-m\right| - \zeta\right) / \tau_\mathrm{L}^{(0)} \;,
\end{equation}
with $\tau_\mathrm{L}^{(0)} = 2\tau_0/(\chi_T n_\mathrm{f}(0))$ the linearized Langevin time~\cite{onsager1934,pearson1954}. 

\emph{Magnetization dynamics.---}
Conveniently, from a conceptual and observational perspective, changes in magnetization are coupled to the current density of free (mobile) monopoles, $j=\mathrm{\partial}{M} /\mathrm{\partial}t=Q_\mathrm{m}n_\mathrm{f}v/\tilde{V}$, where  $\tilde{V}$ is the volume per site and $v=\kappa_\mathrm{m} F = (Q_\mathrm{m}D/k_\mathrm{B}T) F$ the drift velocity. Thus, 
\begin{equation}\label{eq:mag-lin}
\mathrm{d}m/\mathrm{d}t= (h(t) - m) / \tau_\mathrm{m}\;,
\end{equation}
with $\tau_\mathrm{m}= 9 k_\mathrm{B}T \chi_T \tilde{V} / n_\mathrm{f} a^2 \mu_0 Q^2_\mathrm{m} = (3/2)\tau_0/n_\mathrm{f}$ and where, as advertised, $\tau_\mathrm{m}/\tau_\mathrm{L}^{(0)}=\chi_T/2\gg1$ at low temperatures (in DTO $\chi_T\simeq17.5$ at $T=0.45~\mathrm{K}$). Eq.~(\ref{eq:mag-lin}) implies a susceptibility~\cite{ryzhkin2005, ryzhkin2013}
\begin{equation}\label{eq:susc-lin}
\chi(\omega)=\chi_T/(1-i\omega\tau_\mathrm{m})=\chi_T/(1-(3/2)i\omega\tau_0/n_\mathrm{f})\;, 
\end{equation}
which appears Debye-like, but has the relaxation time depend on monopole density. As $n_\mathrm{f}(t)$ is itself time dependent through Eq.~\ref{eq:chem-lin2}, the magnetization relaxtion rate, at low field, is enhanced by a factor $1 + b\zeta/2$ ,
\begin{equation}\label{eq:mag-lin2}
\mathrm{d}m/\mathrm{d}t= (1 + b\zeta/2)(h(t) - m) / \tau_\mathrm{m}\;.
\end{equation}
Equations (\ref{eq:chem-lin2}) and (\ref{eq:mag-lin2}) form our kinetic model which captures the non-equilibrium dynamics observed in simulations: the non-linear behavior comes dominantly from the absolute value in Eq.~(\ref{eq:chem-lin2}) and the monopole-spin coupling term $b\zeta m/2$ in Eq.~(\ref{eq:mag-lin2}). For quantitative comparisons, we replace $b/2$ with the full expression on the right-hand side of Eq.~(\ref{eq:onsager}) and  $\tau_\mathrm{L}^{(0)}$ with $\tau_\mathrm{L}=\tau_\mathrm{L}^{(0)}n_\mathrm{f}(b)/n_\mathrm{f}(0)$, which restores the field-dependence of $\tau_\mathrm{L}$ predicted in~\cite{onsager1934}. The model does not generally permit a solution in closed form, but it is readily integrated numerically and gives quantitative agreement with our numerical data, as shown in Figs.~\ref{fig:quench},\ref{fig:square}a--b,\ref{fig:sine}a--b!

\emph{Non-linear susceptibility.---}
In spin ice the magnetic response is more readily observable than changes in monopole density. In Fig.~\ref{fig:sine} we show that, as for an electrolyte~\cite{mead1939,pearson1954,eigen1955,persoons1979}, enhanced density is stabilized by harmonic driving $H(t)=H_0\sin(\omega t)$, permitting observation in AC susceptibility experiments. Measurements over a large range of driving field amplitudes should give access to the non-linear response, in the form of a non-linear susceptibility, which is the central object of our experimental proposal. 

This non-linear susceptibility $\chi_{H_0}(\omega)=M(\omega)/H_0$ compares the magnetic response $M(\omega)$ to the amplitude $H_0$ of harmonic driving at the same frequency; $M(\omega)$ is obtained by spectral analysis of $M(t)$ in its periodic steady state (details in Supplemental Material). In the low-field limit, the usual linear susceptibility is recovered. Similarly, the response $M(l\omega)$ at multiples of the base frequency yields further susceptibilities $\chi^{(l)}_{H_0}(\omega)$, i.e.~higher harmonics. Both the magnitude of the field response and the occurence of higher harmonics are characteristic of the Wien effect, as it couples a scalar (density) to the \emph{modulus} of an applied vector field~\cite{pearson1954}. Specifically, only odd higher susceptibilities are visible due to the occurence of even harmonics in density.

Harmonic driving stabilises a plateau of density increase in a frequency window between $\omega_\mathrm{low}\simeq1/\tau_\mathrm{m}$ and $\omega_\mathrm{high}\gg1/\tau_\mathrm{L}$ (Fig.~\ref{fig:sine}c). The process limiting the AC Wien effect at high $\omega$ is thus the establishment of Wien-effect correlations, i.e.~the reorientation of bound pairs along the field direction (on a time scale $\tau_\mathrm{or}$), and not the density relaxation ($\tau_\mathrm{L}$); see Ref.~\cite{giblin2011} for an early treatment of reorientation. 

The Wien effect increase in density translates into an increase in susceptibility as observed in Fig.~\ref{fig:sine}c--e. In the simplest approximation, the susceptibility follows the density as in Ryzhkin's original non-interacting theory~\cite{ryzhkin2005}, i.e.~$\Delta \chi_{H_0}(\omega)/\chi_{0}(\omega) \overset{\omega\gg\tau_\mathrm{m}}{=} \Delta \left<n_\mathrm{f}(H_0)\right>_\mathcal{T}/n_\mathrm{f}(0)$. We compute $\chi_{H_0}(\omega)$ in our simulations and observe that this approach is remarkably successful (Fig.~\ref{fig:sine}f); especially so at frequencies $1/\tau_\mathrm{m}\lesssim\omega\lesssim1/\tau_\mathrm{L}$ (Fig.~\ref{fig:sine}e) where density fully relaxes as the field changes (even from zero field, as in Fig.~\ref{fig:quench}c). Further, a reduction in magnetic response is observed between $1/\tau_\mathrm{L}$ and $1/\tau_\mathrm{or}$ reflecting the increasing fraction of time the monopoles spend establishing the Wien effect rather than magnetizing the system. The Wien effect effectively vanishes for $\omega\gg1/\tau_\mathrm{or}$. 

As there are open questions about the exact nature even of the linear response in DTO at $T<T_\mathrm{f}\simeq0.6\mathrm{K}$, at which dynamics slows beyond laboratory time scales~\cite{snyder2001,snyder2004,slobinsky2010,erfanifam2011,pomaranski2013}, it is convenient that our approach relies only on measuring relative quantities, eliminating many non-universal contributions. As an example, $\tau_\mathrm{m}$ can be fixed from the linear susceptibility $\chi_{0}(\omega)$. Moreover, our theory holds above $T_\mathrm{f}$; the $0.7\;\mathrm{K}$ equivalent of Fig.~\ref{fig:sine}e--f is given in Suppl.~Mat. Finally, we note that our model should contribute to the resolution of open issues concerning the experiment of Ref. \cite{bramwell2009} (Refs.~\cite{dunsiger2011,blundell2012,sala2012,quemerais2012,chang2013}, see Ref.~\cite{nuccio2014} for a summary).

\emph{Conclusions.---}
There exists a time and frequency window where the Wien effect in spin ice is just the same as in a weak electrolyte. It seems hard to imagine that the consequent complex magnetic response could have been predicted from the microscopic spin Hamiltonian without the mapping to the monopole Coulomb gas. Our main proposal for experiment concerns the strong amplitude dependence of a non-linear susceptibility. 

The results are also the most detailed modeling of the AC Wien effect in any material system. They enable its study in an ``electrolyte'' which is perfectly symmetric under the interchange of the sign of the charges, and provide access to more delicate aspects of the second Wien effect such as the reorientational dynamics of bound pairs at high frequencies. 

\begin{acknowledgments}
It is a pleasure to thank J. Bloxsom, L. Bovo, C. Castelnovo, M. Gingras, L. Jaubert, A. Sen, and S. Sondhi for discussions and collaborations on related topics. P.C.W.H.~acknowledges financial support from the Institut Universitaire de France.
\end{acknowledgments}

\bibliography{WESI-final}

\begin{thebibliography}{63}%
\makeatletter
\providecommand \@ifxundefined [1]{%
 \@ifx{#1\undefined}
}%
\providecommand \@ifnum [1]{%
 \ifnum #1\expandafter \@firstoftwo
 \else \expandafter \@secondoftwo
 \fi
}%
\providecommand \@ifx [1]{%
 \ifx #1\expandafter \@firstoftwo
 \else \expandafter \@secondoftwo
 \fi
}%
\providecommand \natexlab [1]{#1}%
\providecommand \enquote  [1]{``#1''}%
\providecommand \bibnamefont  [1]{#1}%
\providecommand \bibfnamefont [1]{#1}%
\providecommand \citenamefont [1]{#1}%
\providecommand \href@noop [0]{\@secondoftwo}%
\providecommand \href [0]{\begingroup \@sanitize@url \@href}%
\providecommand \@href[1]{\@@startlink{#1}\@@href}%
\providecommand \@@href[1]{\endgroup#1\@@endlink}%
\providecommand \@sanitize@url [0]{\catcode `\\12\catcode `\$12\catcode
  `\&12\catcode `\#12\catcode `\^12\catcode `\_12\catcode `\%12\relax}%
\providecommand \@@startlink[1]{}%
\providecommand \@@endlink[0]{}%
\providecommand \url  [0]{\begingroup\@sanitize@url \@url }%
\providecommand \@url [1]{\endgroup\@href {#1}{\urlprefix }}%
\providecommand \urlprefix  [0]{URL }%
\providecommand \Eprint [0]{\href }%
\providecommand \doibase [0]{http://dx.doi.org/}%
\providecommand \selectlanguage [0]{\@gobble}%
\providecommand \bibinfo  [0]{\@secondoftwo}%
\providecommand \bibfield  [0]{\@secondoftwo}%
\providecommand \translation [1]{[#1]}%
\providecommand \BibitemOpen [0]{}%
\providecommand \bibitemStop [0]{}%
\providecommand \bibitemNoStop [0]{.\EOS\space}%
\providecommand \EOS [0]{\spacefactor3000\relax}%
\providecommand \BibitemShut  [1]{\csname bibitem#1\endcsname}%
\let\auto@bib@innerbib\@empty
\bibitem [{\citenamefont {Villain}(1979)}]{villain1979}%
  \BibitemOpen
  \bibfield  {author} {\bibinfo {author} {\bibfnamefont {J.}~\bibnamefont
  {Villain}},\ }\href {\doibase 10.1007/BF01325811} {\bibfield  {journal}
  {\bibinfo  {journal} {Zeitschrift f\"{u}r Physik B Condensed Matter}\
  }\textbf {\bibinfo {volume} {33}},\ \bibinfo {pages} {31} (\bibinfo {year}
  {1979})}\BibitemShut {NoStop}%
\bibitem [{\citenamefont {Lacroix}\ \emph {et~al.}(2011)\citenamefont
  {Lacroix}, \citenamefont {Mendels},\ and\ \citenamefont
  {Mila}}]{lacroix2011}%
  \BibitemOpen
  \bibinfo {editor} {\bibfnamefont {C.}~\bibnamefont {Lacroix}}, \bibinfo
  {editor} {\bibfnamefont {P.}~\bibnamefont {Mendels}}, \ and\ \bibinfo
  {editor} {\bibfnamefont {F.}~\bibnamefont {Mila}},\ eds.,\ \href
  {http://link.springer.com/10.1007/978-3-642-10589-0} {\emph {\bibinfo {title}
  {Introduction to Frustrated Magnetism: Materials, Experiments, Theory}}},\
  \bibinfo {series} {Springer Series in {Solid-State} Sciences}, Vol.\ \bibinfo
  {volume} {164}\ (\bibinfo  {publisher} {Springer Berlin Heidelberg},\
  \bibinfo {address} {Berlin, Heidelberg},\ \bibinfo {year} {2011})\BibitemShut
  {NoStop}%
\bibitem [{\citenamefont {Harris}\ \emph {et~al.}(1997)\citenamefont {Harris},
  \citenamefont {Bramwell}, \citenamefont {{McMorrow}}, \citenamefont
  {Zeiske},\ and\ \citenamefont {Godfrey}}]{harris1997}%
  \BibitemOpen
  \bibfield  {author} {\bibinfo {author} {\bibfnamefont {M.~J.}\ \bibnamefont
  {Harris}}, \bibinfo {author} {\bibfnamefont {S.~T.}\ \bibnamefont
  {Bramwell}}, \bibinfo {author} {\bibfnamefont {D.~F.}\ \bibnamefont
  {{McMorrow}}}, \bibinfo {author} {\bibfnamefont {T.}~\bibnamefont {Zeiske}},
  \ and\ \bibinfo {author} {\bibfnamefont {K.~W.}\ \bibnamefont {Godfrey}},\
  }\href {\doibase 10.1103/PhysRevLett.79.2554} {\bibfield  {journal} {\bibinfo
   {journal} {Physical Review Letters}\ }\textbf {\bibinfo {volume} {79}},\
  \bibinfo {pages} {2554} (\bibinfo {year} {1997})}\BibitemShut {NoStop}%
\bibitem [{\citenamefont {Bramwell}\ and\ \citenamefont
  {Gingras}(2001)}]{bramwell2001}%
  \BibitemOpen
  \bibfield  {author} {\bibinfo {author} {\bibfnamefont {S.~T.}\ \bibnamefont
  {Bramwell}}\ and\ \bibinfo {author} {\bibfnamefont {M.~J.~P.}\ \bibnamefont
  {Gingras}},\ }\href {\doibase 10.1126/science.1064761} {\bibfield  {journal}
  {\bibinfo  {journal} {Science}\ }\textbf {\bibinfo {volume} {294}},\ \bibinfo
  {pages} {1495} (\bibinfo {year} {2001})}\BibitemShut {NoStop}%
\bibitem [{\citenamefont {Castelnovo}\ \emph {et~al.}(2012)\citenamefont
  {Castelnovo}, \citenamefont {Moessner},\ and\ \citenamefont
  {Sondhi}}]{castelnovo2012}%
  \BibitemOpen
  \bibfield  {author} {\bibinfo {author} {\bibfnamefont {C.}~\bibnamefont
  {Castelnovo}}, \bibinfo {author} {\bibfnamefont {R.}~\bibnamefont
  {Moessner}}, \ and\ \bibinfo {author} {\bibfnamefont {S.}~\bibnamefont
  {Sondhi}},\ }\href {\doibase 10.1146/annurev-conmatphys-020911-125058}
  {\bibfield  {journal} {\bibinfo  {journal} {Annual Review of Condensed Matter
  Physics}\ }\textbf {\bibinfo {volume} {3}},\ \bibinfo {pages} {35} (\bibinfo
  {year} {2012})}\BibitemShut {NoStop}%
\bibitem [{\citenamefont {Castelnovo}\ \emph {et~al.}(2008)\citenamefont
  {Castelnovo}, \citenamefont {Moessner},\ and\ \citenamefont
  {Sondhi}}]{castelnovo2008}%
  \BibitemOpen
  \bibfield  {author} {\bibinfo {author} {\bibfnamefont {C.}~\bibnamefont
  {Castelnovo}}, \bibinfo {author} {\bibfnamefont {R.}~\bibnamefont
  {Moessner}}, \ and\ \bibinfo {author} {\bibfnamefont {S.~L.}\ \bibnamefont
  {Sondhi}},\ }\href {\doibase 10.1038/nature06433} {\bibfield  {journal}
  {\bibinfo  {journal} {Nature}\ }\textbf {\bibinfo {volume} {451}},\ \bibinfo
  {pages} {42} (\bibinfo {year} {2008})}\BibitemShut {NoStop}%
\bibitem [{\citenamefont {Morris}\ \emph {et~al.}(2009)\citenamefont {Morris},
  \citenamefont {Tennant}, \citenamefont {Grigera}, \citenamefont {Klemke},
  \citenamefont {Castelnovo}, \citenamefont {Moessner}, \citenamefont
  {Czternasty}, \citenamefont {Meissner}, \citenamefont {Rule}, \citenamefont
  {Hoffmann}, \citenamefont {Kiefer}, \citenamefont {Gerischer}, \citenamefont
  {Slobinsky},\ and\ \citenamefont {Perry}}]{morris2009}%
  \BibitemOpen
  \bibfield  {author} {\bibinfo {author} {\bibfnamefont {D.~J.~P.}\
  \bibnamefont {Morris}}, \bibinfo {author} {\bibfnamefont {D.~A.}\
  \bibnamefont {Tennant}}, \bibinfo {author} {\bibfnamefont {S.~A.}\
  \bibnamefont {Grigera}}, \bibinfo {author} {\bibfnamefont {B.}~\bibnamefont
  {Klemke}}, \bibinfo {author} {\bibfnamefont {C.}~\bibnamefont {Castelnovo}},
  \bibinfo {author} {\bibfnamefont {R.}~\bibnamefont {Moessner}}, \bibinfo
  {author} {\bibfnamefont {C.}~\bibnamefont {Czternasty}}, \bibinfo {author}
  {\bibfnamefont {M.}~\bibnamefont {Meissner}}, \bibinfo {author}
  {\bibfnamefont {K.~C.}\ \bibnamefont {Rule}}, \bibinfo {author}
  {\bibfnamefont {J.}~\bibnamefont {Hoffmann}}, \bibinfo {author}
  {\bibfnamefont {K.}~\bibnamefont {Kiefer}}, \bibinfo {author} {\bibfnamefont
  {S.}~\bibnamefont {Gerischer}}, \bibinfo {author} {\bibfnamefont
  {D.}~\bibnamefont {Slobinsky}}, \ and\ \bibinfo {author} {\bibfnamefont
  {R.~S.}\ \bibnamefont {Perry}},\ }\href {\doibase 10.1126/science.1178868}
  {\bibfield  {journal} {\bibinfo  {journal} {Science}\ }\textbf {\bibinfo
  {volume} {326}},\ \bibinfo {pages} {411} (\bibinfo {year}
  {2009})}\BibitemShut {NoStop}%
\bibitem [{\citenamefont {Jaubert}\ and\ \citenamefont
  {Holdsworth}(2011)}]{jaubert2011-a}%
  \BibitemOpen
  \bibfield  {author} {\bibinfo {author} {\bibfnamefont {L.~D.~C.}\
  \bibnamefont {Jaubert}}\ and\ \bibinfo {author} {\bibfnamefont {P.~C.~W.}\
  \bibnamefont {Holdsworth}},\ }\href {\doibase 10.1088/0953-8984/23/16/164222}
  {\bibfield  {journal} {\bibinfo  {journal} {Journal of Physics: Condensed
  Matter}\ }\textbf {\bibinfo {volume} {23}},\ \bibinfo {pages} {164222}
  (\bibinfo {year} {2011})}\BibitemShut {NoStop}%
\bibitem [{\citenamefont {Castelnovo}\ \emph {et~al.}(2011)\citenamefont
  {Castelnovo}, \citenamefont {Moessner},\ and\ \citenamefont
  {Sondhi}}]{castelnovo2011}%
  \BibitemOpen
  \bibfield  {author} {\bibinfo {author} {\bibfnamefont {C.}~\bibnamefont
  {Castelnovo}}, \bibinfo {author} {\bibfnamefont {R.}~\bibnamefont
  {Moessner}}, \ and\ \bibinfo {author} {\bibfnamefont {S.~L.}\ \bibnamefont
  {Sondhi}},\ }\href {\doibase 10.1103/PhysRevB.84.144435} {\bibfield
  {journal} {\bibinfo  {journal} {Physical Review B}\ }\textbf {\bibinfo
  {volume} {84}},\ \bibinfo {pages} {144435} (\bibinfo {year}
  {2011})}\BibitemShut {NoStop}%
\bibitem [{\citenamefont {Onsager}(1934)}]{onsager1934}%
  \BibitemOpen
  \bibfield  {author} {\bibinfo {author} {\bibfnamefont {L.}~\bibnamefont
  {Onsager}},\ }\href {\doibase 10.1063/1.1749541} {\bibfield  {journal}
  {\bibinfo  {journal} {The Journal of Chemical Physics}\ }\textbf {\bibinfo
  {volume} {2}},\ \bibinfo {pages} {599} (\bibinfo {year} {1934})}\BibitemShut
  {NoStop}%
\bibitem [{\citenamefont {Boyd}(2003)}]{boyd2003}%
  \BibitemOpen
  \bibfield  {author} {\bibinfo {author} {\bibfnamefont {R.~W.}\ \bibnamefont
  {Boyd}},\ }\href@noop {} {\emph {\bibinfo {title} {Nonlinear Optics}}}\
  (\bibinfo  {publisher} {Academic Press},\ \bibinfo {year} {2003})\BibitemShut
  {NoStop}%
\bibitem [{\citenamefont {Fujiki}\ and\ \citenamefont
  {Katsura}(1981)}]{fujiki1981}%
  \BibitemOpen
  \bibfield  {author} {\bibinfo {author} {\bibfnamefont {S.}~\bibnamefont
  {Fujiki}}\ and\ \bibinfo {author} {\bibfnamefont {S.}~\bibnamefont
  {Katsura}},\ }\href {\doibase 10.1143/PTP.65.1130} {\bibfield  {journal}
  {\bibinfo  {journal} {Progress of Theoretical Physics}\ }\textbf {\bibinfo
  {volume} {65}},\ \bibinfo {pages} {1130} (\bibinfo {year}
  {1981})}\BibitemShut {NoStop}%
\bibitem [{\citenamefont {Ogielski}(1985)}]{ogielski1985}%
  \BibitemOpen
  \bibfield  {author} {\bibinfo {author} {\bibfnamefont {A.~T.}\ \bibnamefont
  {Ogielski}},\ }\href {\doibase 10.1103/PhysRevB.32.7384} {\bibfield
  {journal} {\bibinfo  {journal} {Physical Review B}\ }\textbf {\bibinfo
  {volume} {32}},\ \bibinfo {pages} {7384} (\bibinfo {year}
  {1985})}\BibitemShut {NoStop}%
\bibitem [{\citenamefont {M\'{e}zard}\ and\ \citenamefont
  {Parisi}(1991)}]{mezard1991}%
  \BibitemOpen
  \bibfield  {author} {\bibinfo {author} {\bibfnamefont {M.}~\bibnamefont
  {M\'{e}zard}}\ and\ \bibinfo {author} {\bibfnamefont {G.}~\bibnamefont
  {Parisi}},\ }\href {\doibase 10.1051/jp1:1991171} {\bibfield  {journal}
  {\bibinfo  {journal} {Journal de Physique I}\ }\textbf {\bibinfo {volume}
  {1}},\ \bibinfo {pages} {809} (\bibinfo {year} {1991})}\BibitemShut {NoStop}%
\bibitem [{\citenamefont {Cugliandolo}\ \emph {et~al.}(1997)\citenamefont
  {Cugliandolo}, \citenamefont {Kurchan},\ and\ \citenamefont
  {Peliti}}]{cugliandolo1997}%
  \BibitemOpen
  \bibfield  {author} {\bibinfo {author} {\bibfnamefont {L.~F.}\ \bibnamefont
  {Cugliandolo}}, \bibinfo {author} {\bibfnamefont {J.}~\bibnamefont
  {Kurchan}}, \ and\ \bibinfo {author} {\bibfnamefont {L.}~\bibnamefont
  {Peliti}},\ }\href {\doibase 10.1103/PhysRevE.55.3898} {\bibfield  {journal}
  {\bibinfo  {journal} {Physical Review E}\ }\textbf {\bibinfo {volume} {55}},\
  \bibinfo {pages} {3898} (\bibinfo {year} {1997})}\BibitemShut {NoStop}%
\bibitem [{\citenamefont {Berthier}\ and\ \citenamefont
  {Biroli}(2011)}]{berthier2011}%
  \BibitemOpen
  \bibfield  {author} {\bibinfo {author} {\bibfnamefont {L.}~\bibnamefont
  {Berthier}}\ and\ \bibinfo {author} {\bibfnamefont {G.}~\bibnamefont
  {Biroli}},\ }\href {\doibase 10.1103/RevModPhys.83.587} {\bibfield  {journal}
  {\bibinfo  {journal} {Reviews of Modern Physics}\ }\textbf {\bibinfo {volume}
  {83}},\ \bibinfo {pages} {587} (\bibinfo {year} {2011})}\BibitemShut
  {NoStop}%
\bibitem [{\citenamefont {Rosenstein}\ and\ \citenamefont
  {Li}(2010)}]{rosenstein2010}%
  \BibitemOpen
  \bibfield  {author} {\bibinfo {author} {\bibfnamefont {B.}~\bibnamefont
  {Rosenstein}}\ and\ \bibinfo {author} {\bibfnamefont {D.}~\bibnamefont
  {Li}},\ }\href {\doibase 10.1103/RevModPhys.82.109} {\bibfield  {journal}
  {\bibinfo  {journal} {Reviews of Modern Physics}\ }\textbf {\bibinfo {volume}
  {82}},\ \bibinfo {pages} {109} (\bibinfo {year} {2010})}\BibitemShut
  {NoStop}%
\bibitem [{\citenamefont {Chandra}\ \emph {et~al.}(2013)\citenamefont
  {Chandra}, \citenamefont {Coleman},\ and\ \citenamefont
  {Flint}}]{chandra2013}%
  \BibitemOpen
  \bibfield  {author} {\bibinfo {author} {\bibfnamefont {P.}~\bibnamefont
  {Chandra}}, \bibinfo {author} {\bibfnamefont {P.}~\bibnamefont {Coleman}}, \
  and\ \bibinfo {author} {\bibfnamefont {R.}~\bibnamefont {Flint}},\ }\href
  {\doibase 10.1038/nature11820} {\bibfield  {journal} {\bibinfo  {journal}
  {Nature}\ }\textbf {\bibinfo {volume} {493}},\ \bibinfo {pages} {621}
  (\bibinfo {year} {2013})}\BibitemShut {NoStop}%
\bibitem [{\citenamefont {Wu}\ \emph {et~al.}(1993)\citenamefont {Wu},
  \citenamefont {Bitko}, \citenamefont {Rosenbaum},\ and\ \citenamefont
  {Aeppli}}]{wu1993}%
  \BibitemOpen
  \bibfield  {author} {\bibinfo {author} {\bibfnamefont {W.}~\bibnamefont
  {Wu}}, \bibinfo {author} {\bibfnamefont {D.}~\bibnamefont {Bitko}}, \bibinfo
  {author} {\bibfnamefont {T.~F.}\ \bibnamefont {Rosenbaum}}, \ and\ \bibinfo
  {author} {\bibfnamefont {G.}~\bibnamefont {Aeppli}},\ }\href {\doibase
  10.1103/PhysRevLett.71.1919} {\bibfield  {journal} {\bibinfo  {journal}
  {Physical Review Letters}\ }\textbf {\bibinfo {volume} {71}},\ \bibinfo
  {pages} {1919} (\bibinfo {year} {1993})}\BibitemShut {NoStop}%
\bibitem [{\citenamefont {Gingras}\ \emph {et~al.}(1997)\citenamefont
  {Gingras}, \citenamefont {Stager}, \citenamefont {Raju}, \citenamefont
  {Gaulin},\ and\ \citenamefont {Greedan}}]{gingras1997}%
  \BibitemOpen
  \bibfield  {author} {\bibinfo {author} {\bibfnamefont {M.~J.~P.}\
  \bibnamefont {Gingras}}, \bibinfo {author} {\bibfnamefont {C.~V.}\
  \bibnamefont {Stager}}, \bibinfo {author} {\bibfnamefont {N.~P.}\
  \bibnamefont {Raju}}, \bibinfo {author} {\bibfnamefont {B.~D.}\ \bibnamefont
  {Gaulin}}, \ and\ \bibinfo {author} {\bibfnamefont {J.~E.}\ \bibnamefont
  {Greedan}},\ }\href {\doibase 10.1103/PhysRevLett.78.947} {\bibfield
  {journal} {\bibinfo  {journal} {Physical Review Letters}\ }\textbf {\bibinfo
  {volume} {78}},\ \bibinfo {pages} {947} (\bibinfo {year} {1997})}\BibitemShut
  {NoStop}%
\bibitem [{\citenamefont {Kaiser}(2014)}]{kaiser2014}%
  \BibitemOpen
  \bibfield  {author} {\bibinfo {author} {\bibfnamefont {V.}~\bibnamefont
  {Kaiser}},\ }\emph {\bibinfo {title} {The Wien Effect in Electric and
  Magnetic Coulomb systems: From Electrolytes to Spin Ice}},\ \href@noop {}
  {\bibinfo {type} {{Ph.D.~Thesis}}},\ \bibinfo  {school} {ENS Lyon / TU
  Dresden} (\bibinfo {year} {2014})\BibitemShut {NoStop}%
\bibitem [{\citenamefont {Ryzhkin}(2005)}]{ryzhkin2005}%
  \BibitemOpen
  \bibfield  {author} {\bibinfo {author} {\bibfnamefont {I.~A.}\ \bibnamefont
  {Ryzhkin}},\ }\href {http://link.springer.com/article/10.1134/1.2103216}
  {\bibfield  {journal} {\bibinfo  {journal} {Journal of Experimental and
  Theoretical Physics}\ }\textbf {\bibinfo {volume} {101}},\ \bibinfo {pages}
  {481{\textendash}486} (\bibinfo {year} {2005})}\BibitemShut {NoStop}%
\bibitem [{\citenamefont {Moessner}\ and\ \citenamefont
  {Sondhi}(2010)}]{moessner2010}%
  \BibitemOpen
  \bibfield  {author} {\bibinfo {author} {\bibfnamefont {R.}~\bibnamefont
  {Moessner}}\ and\ \bibinfo {author} {\bibfnamefont {S.~L.}\ \bibnamefont
  {Sondhi}},\ }\href {\doibase 10.1103/PhysRevLett.105.166401} {\bibfield
  {journal} {\bibinfo  {journal} {Physical Review Letters}\ }\textbf {\bibinfo
  {volume} {105}},\ \bibinfo {pages} {166401} (\bibinfo {year}
  {2010})}\BibitemShut {NoStop}%
\bibitem [{\citenamefont {Fennell}\ \emph {et~al.}(2009)\citenamefont
  {Fennell}, \citenamefont {Deen}, \citenamefont {Wildes}, \citenamefont
  {Schmalzl}, \citenamefont {Prabhakaran}, \citenamefont {Boothroyd},
  \citenamefont {Aldus}, \citenamefont {McMorrow},\ and\ \citenamefont
  {Bramwell}}]{fennell2009}%
  \BibitemOpen
  \bibfield  {author} {\bibinfo {author} {\bibfnamefont {T.}~\bibnamefont
  {Fennell}}, \bibinfo {author} {\bibfnamefont {P.~P.}\ \bibnamefont {Deen}},
  \bibinfo {author} {\bibfnamefont {A.~R.}\ \bibnamefont {Wildes}}, \bibinfo
  {author} {\bibfnamefont {K.}~\bibnamefont {Schmalzl}}, \bibinfo {author}
  {\bibfnamefont {D.}~\bibnamefont {Prabhakaran}}, \bibinfo {author}
  {\bibfnamefont {A.~T.}\ \bibnamefont {Boothroyd}}, \bibinfo {author}
  {\bibfnamefont {R.~J.}\ \bibnamefont {Aldus}}, \bibinfo {author}
  {\bibfnamefont {D.~F.}\ \bibnamefont {McMorrow}}, \ and\ \bibinfo {author}
  {\bibfnamefont {S.~T.}\ \bibnamefont {Bramwell}},\ }\href
  {http://www.sciencemag.org/content/326/5951/415.short} {\bibfield  {journal}
  {\bibinfo  {journal} {Science}\ }\textbf {\bibinfo {volume} {326}},\ \bibinfo
  {pages} {415} (\bibinfo {year} {2009})}\BibitemShut {NoStop}%
\bibitem [{\citenamefont {Jaubert}\ and\ \citenamefont
  {Holdsworth}(2009)}]{jaubert2009}%
  \BibitemOpen
  \bibfield  {author} {\bibinfo {author} {\bibfnamefont {L.~D.~C.}\
  \bibnamefont {Jaubert}}\ and\ \bibinfo {author} {\bibfnamefont {P.~C.~W.}\
  \bibnamefont {Holdsworth}},\ }\href {\doibase 10.1038/nphys1227} {\bibfield
  {journal} {\bibinfo  {journal} {Nature Physics}\ }\textbf {\bibinfo {volume}
  {5}},\ \bibinfo {pages} {258} (\bibinfo {year} {2009})}\BibitemShut {NoStop}%
\bibitem [{\citenamefont {de~Leeuw}\ \emph {et~al.}(1980)\citenamefont
  {de~Leeuw}, \citenamefont {Perram},\ and\ \citenamefont
  {Smith}}]{deleeuw1980}%
  \BibitemOpen
  \bibfield  {author} {\bibinfo {author} {\bibfnamefont {S.~W.}\ \bibnamefont
  {de~Leeuw}}, \bibinfo {author} {\bibfnamefont {J.~W.}\ \bibnamefont
  {Perram}}, \ and\ \bibinfo {author} {\bibfnamefont {E.~R.}\ \bibnamefont
  {Smith}},\ }\href
  {http://rspa.royalsocietypublishing.org/content/373/1752/27.short} {\bibfield
   {journal} {\bibinfo  {journal} {Proceedings of the Royal Society of London.
  A. Mathematical and Physical Sciences}\ }\textbf {\bibinfo {volume} {373}},\
  \bibinfo {pages} {27{\textendash}56} (\bibinfo {year} {1980})}\BibitemShut
  {NoStop}%
\bibitem [{\citenamefont {Barkema}\ and\ \citenamefont
  {Newman}(1998)}]{barkema1998}%
  \BibitemOpen
  \bibfield  {author} {\bibinfo {author} {\bibfnamefont {G.~T.}\ \bibnamefont
  {Barkema}}\ and\ \bibinfo {author} {\bibfnamefont {M.~E.~J.}\ \bibnamefont
  {Newman}},\ }\href
  {http://journals.aps.org/pre/abstract/10.1103/PhysRevE.57.1155} {\bibfield
  {journal} {\bibinfo  {journal} {Physical Review E}\ }\textbf {\bibinfo
  {volume} {57}},\ \bibinfo {pages} {1155} (\bibinfo {year}
  {1998})}\BibitemShut {NoStop}%
\bibitem [{\citenamefont {Melko}\ and\ \citenamefont
  {Gingras}(2004)}]{melko2004}%
  \BibitemOpen
  \bibfield  {author} {\bibinfo {author} {\bibfnamefont {R.~G.}\ \bibnamefont
  {Melko}}\ and\ \bibinfo {author} {\bibfnamefont {M.~J.~P.}\ \bibnamefont
  {Gingras}},\ }\href {\doibase 10.1088/0953-8984/16/43/R02} {\bibfield
  {journal} {\bibinfo  {journal} {Journal of Physics: Condensed Matter}\
  }\textbf {\bibinfo {volume} {16}},\ \bibinfo {pages} {R1277} (\bibinfo {year}
  {2004})}\BibitemShut {NoStop}%
\bibitem [{\citenamefont {Jaubert}\ \emph {et~al.}(2011)\citenamefont
  {Jaubert}, \citenamefont {Haque},\ and\ \citenamefont
  {Moessner}}]{jaubert2011}%
  \BibitemOpen
  \bibfield  {author} {\bibinfo {author} {\bibfnamefont {L.~D.~C.}\
  \bibnamefont {Jaubert}}, \bibinfo {author} {\bibfnamefont {M.}~\bibnamefont
  {Haque}}, \ and\ \bibinfo {author} {\bibfnamefont {R.}~\bibnamefont
  {Moessner}},\ }\href
  {http://journals.aps.org/prl/abstract/10.1103/PhysRevLett.107.177202}
  {\bibfield  {journal} {\bibinfo  {journal} {Physical review letters}\
  }\textbf {\bibinfo {volume} {107}},\ \bibinfo {pages} {177202} (\bibinfo
  {year} {2011})}\BibitemShut {NoStop}%
\bibitem [{\citenamefont {Kikuchi}\ \emph {et~al.}(1991)\citenamefont
  {Kikuchi}, \citenamefont {Yoshida}, \citenamefont {Maekawa},\ and\
  \citenamefont {Watanabe}}]{kikuchi1991}%
  \BibitemOpen
  \bibfield  {author} {\bibinfo {author} {\bibfnamefont {K.}~\bibnamefont
  {Kikuchi}}, \bibinfo {author} {\bibfnamefont {M.}~\bibnamefont {Yoshida}},
  \bibinfo {author} {\bibfnamefont {T.}~\bibnamefont {Maekawa}}, \ and\
  \bibinfo {author} {\bibfnamefont {H.}~\bibnamefont {Watanabe}},\ }\href
  {\doibase 10.1016/S0009-2614(91)85070-D} {\bibfield  {journal} {\bibinfo
  {journal} {Chemical Physics Letters}\ }\textbf {\bibinfo {volume} {185}},\
  \bibinfo {pages} {335} (\bibinfo {year} {1991})}\BibitemShut {NoStop}%
\bibitem [{\citenamefont {Cheng}\ \emph {et~al.}(2006)\citenamefont {Cheng},
  \citenamefont {Jalil}, \citenamefont {Lee},\ and\ \citenamefont
  {Okabe}}]{cheng2006}%
  \BibitemOpen
  \bibfield  {author} {\bibinfo {author} {\bibfnamefont {X.~Z.}\ \bibnamefont
  {Cheng}}, \bibinfo {author} {\bibfnamefont {M.~B.~A.}\ \bibnamefont {Jalil}},
  \bibinfo {author} {\bibfnamefont {H.~K.}\ \bibnamefont {Lee}}, \ and\
  \bibinfo {author} {\bibfnamefont {Y.}~\bibnamefont {Okabe}},\ }\href
  {\doibase 10.1103/PhysRevLett.96.067208} {\bibfield  {journal} {\bibinfo
  {journal} {Physical Review Letters}\ }\textbf {\bibinfo {volume} {96}},\
  \bibinfo {pages} {067208} (\bibinfo {year} {2006})}\BibitemShut {NoStop}%
\bibitem [{\citenamefont {Sanz}\ and\ \citenamefont
  {Marenduzzo}(2010)}]{sanz2010}%
  \BibitemOpen
  \bibfield  {author} {\bibinfo {author} {\bibfnamefont {E.}~\bibnamefont
  {Sanz}}\ and\ \bibinfo {author} {\bibfnamefont {D.}~\bibnamefont
  {Marenduzzo}},\ }\href {\doibase 10.1063/1.3414827} {\bibfield  {journal}
  {\bibinfo  {journal} {The Journal of Chemical Physics}\ }\textbf {\bibinfo
  {volume} {132}},\ \bibinfo {pages} {194102} (\bibinfo {year}
  {2010})}\BibitemShut {NoStop}%
\bibitem [{\citenamefont {Bramwell}\ \emph {et~al.}(2009)\citenamefont
  {Bramwell}, \citenamefont {Giblin}, \citenamefont {Calder}, \citenamefont
  {Aldus}, \citenamefont {Prabhakaran},\ and\ \citenamefont
  {Fennell}}]{bramwell2009}%
  \BibitemOpen
  \bibfield  {author} {\bibinfo {author} {\bibfnamefont {S.~T.}\ \bibnamefont
  {Bramwell}}, \bibinfo {author} {\bibfnamefont {S.~R.}\ \bibnamefont
  {Giblin}}, \bibinfo {author} {\bibfnamefont {S.}~\bibnamefont {Calder}},
  \bibinfo {author} {\bibfnamefont {R.}~\bibnamefont {Aldus}}, \bibinfo
  {author} {\bibfnamefont {D.}~\bibnamefont {Prabhakaran}}, \ and\ \bibinfo
  {author} {\bibfnamefont {T.}~\bibnamefont {Fennell}},\ }\href {\doibase
  10.1038/nature08500} {\bibfield  {journal} {\bibinfo  {journal} {Nature}\
  }\textbf {\bibinfo {volume} {461}},\ \bibinfo {pages} {956} (\bibinfo {year}
  {2009})}\BibitemShut {NoStop}%
\bibitem [{\citenamefont {Bjerrum}(1926)}]{bjerrum1926}%
  \BibitemOpen
  \bibfield  {author} {\bibinfo {author} {\bibfnamefont {N.~J.}\ \bibnamefont
  {Bjerrum}},\ }\href
  {http://www.sdu.dk/media/bibpdf/Bind%201-9%5CBind%5Cmfm-7-9.pdf} {\bibfield
  {journal} {\bibinfo  {journal} {Det Kgl. Danske Videnskabernes Selskab.,
  Mathematisk-fysiske Meddelelser.}\ }\textbf {\bibinfo {volume} {7}},\
  \bibinfo {pages} {1} (\bibinfo {year} {1926})}\BibitemShut {NoStop}%
\bibitem [{\citenamefont {Kobelev}\ \emph {et~al.}(2002)\citenamefont
  {Kobelev}, \citenamefont {Kolomeisky},\ and\ \citenamefont
  {Fisher}}]{kobelev2002}%
  \BibitemOpen
  \bibfield  {author} {\bibinfo {author} {\bibfnamefont {V.}~\bibnamefont
  {Kobelev}}, \bibinfo {author} {\bibfnamefont {A.~B.}\ \bibnamefont
  {Kolomeisky}}, \ and\ \bibinfo {author} {\bibfnamefont {M.~E.}\ \bibnamefont
  {Fisher}},\ }\href {\doibase 10.1063/1.1464827} {\bibfield  {journal}
  {\bibinfo  {journal} {The Journal of Chemical Physics}\ }\textbf {\bibinfo
  {volume} {116}},\ \bibinfo {pages} {7589} (\bibinfo {year}
  {2002})}\BibitemShut {NoStop}%
\bibitem [{\citenamefont {Persoons}\ and\ \citenamefont
  {Beylen}(1979)}]{persoons1979}%
  \BibitemOpen
  \bibfield  {author} {\bibinfo {author} {\bibfnamefont {A.}~\bibnamefont
  {Persoons}}\ and\ \bibinfo {author} {\bibfnamefont {M.~V.}\ \bibnamefont
  {Beylen}},\ }\href {\doibase 10.1351/pac197951040887} {\bibfield  {journal}
  {\bibinfo  {journal} {Pure and Applied Chemistry}\ }\textbf {\bibinfo
  {volume} {51}},\ \bibinfo {pages} {887{\textendash}900} (\bibinfo {year}
  {1979})}\BibitemShut {NoStop}%
\bibitem [{\citenamefont {Debye}\ and\ \citenamefont
  {H{\"u}ckel}(1923)}]{debye1923}%
  \BibitemOpen
  \bibfield  {author} {\bibinfo {author} {\bibfnamefont {P.}~\bibnamefont
  {Debye}}\ and\ \bibinfo {author} {\bibfnamefont {E.}~\bibnamefont
  {H{\"u}ckel}},\ }\href@noop {} {\bibfield  {journal} {\bibinfo  {journal}
  {Physikalische Zeitschrift}\ }\textbf {\bibinfo {volume} {24}},\ \bibinfo
  {pages} {185} (\bibinfo {year} {1923})}\BibitemShut {NoStop}%
\bibitem [{\citenamefont {Moore}(1999)}]{moore1999}%
  \BibitemOpen
  \bibfield  {author} {\bibinfo {author} {\bibfnamefont {W.~J.}\ \bibnamefont
  {Moore}},\ }\href@noop {} {\emph {\bibinfo {title} {Physical Chemistry}}},\
  \bibinfo {edition} {5th}\ ed.\ (\bibinfo  {publisher} {{Prentice-Hall}},\
  \bibinfo {year} {1999})\BibitemShut {NoStop}%
\bibitem [{\citenamefont {Langevin}(1903)}]{langevin1903}%
  \BibitemOpen
  \bibfield  {author} {\bibinfo {author} {\bibfnamefont {P.}~\bibnamefont
  {Langevin}},\ }\href@noop {} {\bibfield  {journal} {\bibinfo  {journal}
  {Annales de chimie et de physique}\ }\bibinfo {series} {7},\ \textbf
  {\bibinfo {volume} {28}},\ \bibinfo {pages} {433} (\bibinfo {year}
  {1903})}\BibitemShut {NoStop}%
\bibitem [{\citenamefont {Jaubert}\ \emph {et~al.}(2013)\citenamefont
  {Jaubert}, \citenamefont {Harris}, \citenamefont {Fennell}, \citenamefont
  {Melko}, \citenamefont {Bramwell},\ and\ \citenamefont
  {Holdsworth}}]{jaubert2013}%
  \BibitemOpen
  \bibfield  {author} {\bibinfo {author} {\bibfnamefont {L.}~\bibnamefont
  {Jaubert}}, \bibinfo {author} {\bibfnamefont {M.}~\bibnamefont {Harris}},
  \bibinfo {author} {\bibfnamefont {T.}~\bibnamefont {Fennell}}, \bibinfo
  {author} {\bibfnamefont {R.}~\bibnamefont {Melko}}, \bibinfo {author}
  {\bibfnamefont {S.}~\bibnamefont {Bramwell}}, \ and\ \bibinfo {author}
  {\bibfnamefont {P.}~\bibnamefont {Holdsworth}},\ }\href {\doibase
  10.1103/PhysRevX.3.011014} {\bibfield  {journal} {\bibinfo  {journal}
  {Physical Review X}\ }\textbf {\bibinfo {volume} {3}},\ \bibinfo {pages}
  {011014} (\bibinfo {year} {2013})}\BibitemShut {NoStop}%
\bibitem [{\citenamefont {Onsager}\ and\ \citenamefont
  {Kim}(1957)}]{onsager1957}%
  \BibitemOpen
  \bibfield  {author} {\bibinfo {author} {\bibfnamefont {L.}~\bibnamefont
  {Onsager}}\ and\ \bibinfo {author} {\bibfnamefont {S.~K.}\ \bibnamefont
  {Kim}},\ }\href {\doibase 10.1021/j150548a015} {\bibfield  {journal}
  {\bibinfo  {journal} {The Journal of Physical Chemistry}\ }\textbf {\bibinfo
  {volume} {61}},\ \bibinfo {pages} {198} (\bibinfo {year} {1957})}\BibitemShut
  {NoStop}%
\bibitem [{\citenamefont {Patterson}\ and\ \citenamefont
  {Freitag}(1961)}]{patterson1961}%
  \BibitemOpen
  \bibfield  {author} {\bibinfo {author} {\bibfnamefont {A.}~\bibnamefont
  {Patterson}}\ and\ \bibinfo {author} {\bibfnamefont {H.}~\bibnamefont
  {Freitag}},\ }\href {\doibase 10.1149/1.2428129} {\bibfield  {journal}
  {\bibinfo  {journal} {Journal of The Electrochemical Society}\ }\textbf
  {\bibinfo {volume} {108}},\ \bibinfo {pages} {529} (\bibinfo {year}
  {1961})}\BibitemShut {NoStop}%
\bibitem [{\citenamefont {Kaiser}\ \emph {et~al.}(2013)\citenamefont {Kaiser},
  \citenamefont {Bramwell}, \citenamefont {Holdsworth},\ and\ \citenamefont
  {Moessner}}]{kaiser2013}%
  \BibitemOpen
  \bibfield  {author} {\bibinfo {author} {\bibfnamefont {V.}~\bibnamefont
  {Kaiser}}, \bibinfo {author} {\bibfnamefont {S.~T.}\ \bibnamefont
  {Bramwell}}, \bibinfo {author} {\bibfnamefont {P.~C.~W.}\ \bibnamefont
  {Holdsworth}}, \ and\ \bibinfo {author} {\bibfnamefont {R.}~\bibnamefont
  {Moessner}},\ }\href {\doibase 10.1038/nmat3729} {\bibfield  {journal}
  {\bibinfo  {journal} {Nature Materials}\ }\textbf {\bibinfo {volume} {12}},\
  \bibinfo {pages} {1033} (\bibinfo {year} {2013})}\BibitemShut {NoStop}%
\bibitem [{\citenamefont {Pearson}(1954)}]{pearson1954}%
  \BibitemOpen
  \bibfield  {author} {\bibinfo {author} {\bibfnamefont {R.~G.}\ \bibnamefont
  {Pearson}},\ }\href {\doibase 10.1039/df9541700187} {\bibfield  {journal}
  {\bibinfo  {journal} {Discussions of the Faraday Society}\ }\textbf {\bibinfo
  {volume} {17}},\ \bibinfo {pages} {187} (\bibinfo {year} {1954})}\BibitemShut
  {NoStop}%
\bibitem [{\citenamefont {Ryzhkin}\ \emph {et~al.}(2013)\citenamefont
  {Ryzhkin}, \citenamefont {Ryzhkin},\ and\ \citenamefont
  {Bramwell}}]{ryzhkin2013}%
  \BibitemOpen
  \bibfield  {author} {\bibinfo {author} {\bibfnamefont {M.~I.}\ \bibnamefont
  {Ryzhkin}}, \bibinfo {author} {\bibfnamefont {I.~A.}\ \bibnamefont
  {Ryzhkin}}, \ and\ \bibinfo {author} {\bibfnamefont {S.~T.}\ \bibnamefont
  {Bramwell}},\ }\href {\doibase 10.1209/0295-5075/104/37005} {\bibfield
  {journal} {\bibinfo  {journal} {{EPL} {(Europhysics} Letters)}\ }\textbf
  {\bibinfo {volume} {104}},\ \bibinfo {pages} {37005} (\bibinfo {year}
  {2013})}\BibitemShut {NoStop}%
\bibitem [{\citenamefont {Mead}\ and\ \citenamefont {Fuoss}(1939)}]{mead1939}%
  \BibitemOpen
  \bibfield  {author} {\bibinfo {author} {\bibfnamefont {D.~J.}\ \bibnamefont
  {Mead}}\ and\ \bibinfo {author} {\bibfnamefont {R.~M.}\ \bibnamefont
  {Fuoss}},\ }\href {http://pubs.acs.org/doi/abs/10.1021/ja01877a028}
  {\bibfield  {journal} {\bibinfo  {journal} {Journal of the American Chemical
  Society}\ }\textbf {\bibinfo {volume} {61}},\ \bibinfo {pages}
  {2047{\textendash}2053} (\bibinfo {year} {1939})}\BibitemShut {NoStop}%
\bibitem [{\citenamefont {Eigen}\ and\ \citenamefont
  {Schoen}(1955)}]{eigen1955}%
  \BibitemOpen
  \bibfield  {author} {\bibinfo {author} {\bibfnamefont {M.}~\bibnamefont
  {Eigen}}\ and\ \bibinfo {author} {\bibfnamefont {J.}~\bibnamefont {Schoen}},\
  }\href {\doibase 10.1002/bbpc.19550590604} {\bibfield  {journal} {\bibinfo
  {journal} {Zeitschrift f{\"u}r Elektrochemie, Berichte der Bunsengesellschaft
  f{\"u}r physikalische Chemie}\ }\textbf {\bibinfo {volume} {59}},\ \bibinfo
  {pages} {483} (\bibinfo {year} {1955})}\BibitemShut {NoStop}%
\bibitem [{\citenamefont {Giblin}\ \emph {et~al.}(2011)\citenamefont {Giblin},
  \citenamefont {Bramwell}, \citenamefont {Holdsworth}, \citenamefont
  {Prabhakaran},\ and\ \citenamefont {Terry}}]{giblin2011}%
  \BibitemOpen
  \bibfield  {author} {\bibinfo {author} {\bibfnamefont {S.~R.}\ \bibnamefont
  {Giblin}}, \bibinfo {author} {\bibfnamefont {S.~T.}\ \bibnamefont
  {Bramwell}}, \bibinfo {author} {\bibfnamefont {P.~C.~W.}\ \bibnamefont
  {Holdsworth}}, \bibinfo {author} {\bibfnamefont {D.}~\bibnamefont
  {Prabhakaran}}, \ and\ \bibinfo {author} {\bibfnamefont {I.}~\bibnamefont
  {Terry}},\ }\href {\doibase 10.1038/nphys1896} {\bibfield  {journal}
  {\bibinfo  {journal} {Nature Physics}\ }\textbf {\bibinfo {volume} {7}},\
  \bibinfo {pages} {252} (\bibinfo {year} {2011})}\BibitemShut {NoStop}%
\bibitem [{\citenamefont {Snyder}\ \emph {et~al.}(2001)\citenamefont {Snyder},
  \citenamefont {Slusky}, \citenamefont {Cava},\ and\ \citenamefont
  {Schiffer}}]{snyder2001}%
  \BibitemOpen
  \bibfield  {author} {\bibinfo {author} {\bibfnamefont {J.}~\bibnamefont
  {Snyder}}, \bibinfo {author} {\bibfnamefont {J.~S.}\ \bibnamefont {Slusky}},
  \bibinfo {author} {\bibfnamefont {R.~J.}\ \bibnamefont {Cava}}, \ and\
  \bibinfo {author} {\bibfnamefont {P.}~\bibnamefont {Schiffer}},\ }\href
  {\doibase 10.1038/35092516} {\bibfield  {journal} {\bibinfo  {journal}
  {Nature}\ }\textbf {\bibinfo {volume} {413}},\ \bibinfo {pages} {48}
  (\bibinfo {year} {2001})}\BibitemShut {NoStop}%
\bibitem [{\citenamefont {Snyder}\ \emph {et~al.}(2004)\citenamefont {Snyder},
  \citenamefont {Ueland}, \citenamefont {Slusky}, \citenamefont {Karunadasa},
  \citenamefont {Cava},\ and\ \citenamefont {Schiffer}}]{snyder2004}%
  \BibitemOpen
  \bibfield  {author} {\bibinfo {author} {\bibfnamefont {J.}~\bibnamefont
  {Snyder}}, \bibinfo {author} {\bibfnamefont {B.}~\bibnamefont {Ueland}},
  \bibinfo {author} {\bibfnamefont {J.}~\bibnamefont {Slusky}}, \bibinfo
  {author} {\bibfnamefont {H.}~\bibnamefont {Karunadasa}}, \bibinfo {author}
  {\bibfnamefont {R.}~\bibnamefont {Cava}}, \ and\ \bibinfo {author}
  {\bibfnamefont {P.}~\bibnamefont {Schiffer}},\ }\href {\doibase
  10.1103/PhysRevB.69.064414} {\bibfield  {journal} {\bibinfo  {journal}
  {Physical Review B}\ }\textbf {\bibinfo {volume} {69}},\ \bibinfo {pages}
  {064414} (\bibinfo {year} {2004})}\BibitemShut {NoStop}%
\bibitem [{\citenamefont {Slobinsky}\ \emph {et~al.}(2010)\citenamefont
  {Slobinsky}, \citenamefont {Castelnovo}, \citenamefont {Borzi}, \citenamefont
  {Gibbs}, \citenamefont {Mackenzie}, \citenamefont {Moessner},\ and\
  \citenamefont {Grigera}}]{slobinsky2010}%
  \BibitemOpen
  \bibfield  {author} {\bibinfo {author} {\bibfnamefont {D.}~\bibnamefont
  {Slobinsky}}, \bibinfo {author} {\bibfnamefont {C.}~\bibnamefont
  {Castelnovo}}, \bibinfo {author} {\bibfnamefont {R.~A.}\ \bibnamefont
  {Borzi}}, \bibinfo {author} {\bibfnamefont {A.~S.}\ \bibnamefont {Gibbs}},
  \bibinfo {author} {\bibfnamefont {A.~P.}\ \bibnamefont {Mackenzie}}, \bibinfo
  {author} {\bibfnamefont {R.}~\bibnamefont {Moessner}}, \ and\ \bibinfo
  {author} {\bibfnamefont {S.~A.}\ \bibnamefont {Grigera}},\ }\href {\doibase
  10.1103/PhysRevLett.105.267205} {\bibfield  {journal} {\bibinfo  {journal}
  {Physical Review Letters}\ }\textbf {\bibinfo {volume} {105}},\ \bibinfo
  {pages} {267205} (\bibinfo {year} {2010})}\BibitemShut {NoStop}%
\bibitem [{\citenamefont {Erfanifam}\ \emph {et~al.}(2011)\citenamefont
  {Erfanifam}, \citenamefont {Zherlitsyn}, \citenamefont {Wosnitza},
  \citenamefont {Moessner}, \citenamefont {Petrenko}, \citenamefont
  {Balakrishnan},\ and\ \citenamefont {Zvyagin}}]{erfanifam2011}%
  \BibitemOpen
  \bibfield  {author} {\bibinfo {author} {\bibfnamefont {S.}~\bibnamefont
  {Erfanifam}}, \bibinfo {author} {\bibfnamefont {S.}~\bibnamefont
  {Zherlitsyn}}, \bibinfo {author} {\bibfnamefont {J.}~\bibnamefont
  {Wosnitza}}, \bibinfo {author} {\bibfnamefont {R.}~\bibnamefont {Moessner}},
  \bibinfo {author} {\bibfnamefont {O.}~\bibnamefont {Petrenko}}, \bibinfo
  {author} {\bibfnamefont {G.}~\bibnamefont {Balakrishnan}}, \ and\ \bibinfo
  {author} {\bibfnamefont {A.}~\bibnamefont {Zvyagin}},\ }\href {\doibase
  10.1103/PhysRevB.84.220404} {\bibfield  {journal} {\bibinfo  {journal}
  {Physical Review B}\ }\textbf {\bibinfo {volume} {84}},\ \bibinfo {pages}
  {220404} (\bibinfo {year} {2011})}\BibitemShut {NoStop}%
\bibitem [{\citenamefont {Pomaranski}\ \emph {et~al.}(2013)\citenamefont
  {Pomaranski}, \citenamefont {Yaraskavitch}, \citenamefont {Meng},
  \citenamefont {Ross}, \citenamefont {Noad}, \citenamefont {Dabkowska},
  \citenamefont {Gaulin},\ and\ \citenamefont {Kycia}}]{pomaranski2013}%
  \BibitemOpen
  \bibfield  {author} {\bibinfo {author} {\bibfnamefont {D.}~\bibnamefont
  {Pomaranski}}, \bibinfo {author} {\bibfnamefont {L.~R.}\ \bibnamefont
  {Yaraskavitch}}, \bibinfo {author} {\bibfnamefont {S.}~\bibnamefont {Meng}},
  \bibinfo {author} {\bibfnamefont {K.~A.}\ \bibnamefont {Ross}}, \bibinfo
  {author} {\bibfnamefont {H.~M.~L.}\ \bibnamefont {Noad}}, \bibinfo {author}
  {\bibfnamefont {H.~A.}\ \bibnamefont {Dabkowska}}, \bibinfo {author}
  {\bibfnamefont {B.~D.}\ \bibnamefont {Gaulin}}, \ and\ \bibinfo {author}
  {\bibfnamefont {J.~B.}\ \bibnamefont {Kycia}},\ }\href {\doibase
  10.1038/nphys2591} {\bibfield  {journal} {\bibinfo  {journal} {Nature
  Physics}\ }\textbf {\bibinfo {volume} {9}},\ \bibinfo {pages} {353} (\bibinfo
  {year} {2013})}\BibitemShut {NoStop}%
\bibitem [{\citenamefont {Dunsiger}\ \emph {et~al.}(2011)\citenamefont
  {Dunsiger}, \citenamefont {Aczel}, \citenamefont {Arguello}, \citenamefont
  {Dabkowska}, \citenamefont {Dabkowski}, \citenamefont {Du}, \citenamefont
  {Goko}, \citenamefont {Javanparast}, \citenamefont {Lin}, \citenamefont
  {Ning}, \citenamefont {Noad}, \citenamefont {Singh}, \citenamefont
  {Williams}, \citenamefont {Uemura}, \citenamefont {Gingras},\ and\
  \citenamefont {Luke}}]{dunsiger2011}%
  \BibitemOpen
  \bibfield  {author} {\bibinfo {author} {\bibfnamefont {S.}~\bibnamefont
  {Dunsiger}}, \bibinfo {author} {\bibfnamefont {A.}~\bibnamefont {Aczel}},
  \bibinfo {author} {\bibfnamefont {C.}~\bibnamefont {Arguello}}, \bibinfo
  {author} {\bibfnamefont {H.}~\bibnamefont {Dabkowska}}, \bibinfo {author}
  {\bibfnamefont {A.}~\bibnamefont {Dabkowski}}, \bibinfo {author}
  {\bibfnamefont {M.-H.}\ \bibnamefont {Du}}, \bibinfo {author} {\bibfnamefont
  {T.}~\bibnamefont {Goko}}, \bibinfo {author} {\bibfnamefont {B.}~\bibnamefont
  {Javanparast}}, \bibinfo {author} {\bibfnamefont {T.}~\bibnamefont {Lin}},
  \bibinfo {author} {\bibfnamefont {F.}~\bibnamefont {Ning}}, \bibinfo {author}
  {\bibfnamefont {H.}~\bibnamefont {Noad}}, \bibinfo {author} {\bibfnamefont
  {D.}~\bibnamefont {Singh}}, \bibinfo {author} {\bibfnamefont
  {T.}~\bibnamefont {Williams}}, \bibinfo {author} {\bibfnamefont
  {Y.}~\bibnamefont {Uemura}}, \bibinfo {author} {\bibfnamefont
  {M.}~\bibnamefont {Gingras}}, \ and\ \bibinfo {author} {\bibfnamefont
  {G.}~\bibnamefont {Luke}},\ }\href {\doibase 10.1103/PhysRevLett.107.207207}
  {\bibfield  {journal} {\bibinfo  {journal} {Phys. Rev. Lett.}\ }\textbf
  {\bibinfo {volume} {107}},\ \bibinfo {pages} {207207} (\bibinfo {year}
  {2011})}\BibitemShut {NoStop}%
\bibitem [{\citenamefont {Blundell}(2012)}]{blundell2012}%
  \BibitemOpen
  \bibfield  {author} {\bibinfo {author} {\bibfnamefont {S.~J.}\ \bibnamefont
  {Blundell}},\ }\href {\doibase 10.1103/PhysRevLett.108.147601} {\bibfield
  {journal} {\bibinfo  {journal} {Physical Review Letters}\ }\textbf {\bibinfo
  {volume} {108}},\ \bibinfo {pages} {147601} (\bibinfo {year}
  {2012})}\BibitemShut {NoStop}%
\bibitem [{\citenamefont {Sala}\ \emph {et~al.}(2012)\citenamefont {Sala},
  \citenamefont {Castelnovo}, \citenamefont {Moessner}, \citenamefont {Sondhi},
  \citenamefont {Kitagawa}, \citenamefont {Takigawa}, \citenamefont
  {Higashinaka},\ and\ \citenamefont {Maeno}}]{sala2012}%
  \BibitemOpen
  \bibfield  {author} {\bibinfo {author} {\bibfnamefont {G.}~\bibnamefont
  {Sala}}, \bibinfo {author} {\bibfnamefont {C.}~\bibnamefont {Castelnovo}},
  \bibinfo {author} {\bibfnamefont {R.}~\bibnamefont {Moessner}}, \bibinfo
  {author} {\bibfnamefont {S.~L.}\ \bibnamefont {Sondhi}}, \bibinfo {author}
  {\bibfnamefont {K.}~\bibnamefont {Kitagawa}}, \bibinfo {author}
  {\bibfnamefont {M.}~\bibnamefont {Takigawa}}, \bibinfo {author}
  {\bibfnamefont {R.}~\bibnamefont {Higashinaka}}, \ and\ \bibinfo {author}
  {\bibfnamefont {Y.}~\bibnamefont {Maeno}},\ }\href {\doibase
  10.1103/PhysRevLett.108.217203} {\bibfield  {journal} {\bibinfo  {journal}
  {Physical Review Letters}\ }\textbf {\bibinfo {volume} {108}},\ \bibinfo
  {pages} {217203} (\bibinfo {year} {2012})}\BibitemShut {NoStop}%
\bibitem [{\citenamefont {Qu{\'e}merais}\ \emph {et~al.}(2012)\citenamefont
  {Qu{\'e}merais}, \citenamefont {McClarty},\ and\ \citenamefont
  {Moessner}}]{quemerais2012}%
  \BibitemOpen
  \bibfield  {author} {\bibinfo {author} {\bibfnamefont {P.}~\bibnamefont
  {Qu{\'e}merais}}, \bibinfo {author} {\bibfnamefont {P.}~\bibnamefont
  {McClarty}}, \ and\ \bibinfo {author} {\bibfnamefont {R.}~\bibnamefont
  {Moessner}},\ }\href {\doibase 10.1103/PhysRevLett.109.127601} {\bibfield
  {journal} {\bibinfo  {journal} {Physical Review Letters}\ }\textbf {\bibinfo
  {volume} {109}},\ \bibinfo {pages} {127601} (\bibinfo {year}
  {2012})}\BibitemShut {NoStop}%
\bibitem [{\citenamefont {Chang}\ \emph {et~al.}(2013)\citenamefont {Chang},
  \citenamefont {Lees}, \citenamefont {Balakrishnan}, \citenamefont {Kao},\
  and\ \citenamefont {Hillier}}]{chang2013}%
  \BibitemOpen
  \bibfield  {author} {\bibinfo {author} {\bibfnamefont {L.~J.}\ \bibnamefont
  {Chang}}, \bibinfo {author} {\bibfnamefont {M.~R.}\ \bibnamefont {Lees}},
  \bibinfo {author} {\bibfnamefont {G.}~\bibnamefont {Balakrishnan}}, \bibinfo
  {author} {\bibfnamefont {Y.-J.}\ \bibnamefont {Kao}}, \ and\ \bibinfo
  {author} {\bibfnamefont {A.~D.}\ \bibnamefont {Hillier}},\ }\href
  {http://www.nature.com/srep/2013/130523/srep01881/full/srep01881.html?message-global=remove}
  {\bibfield  {journal} {\bibinfo  {journal} {Scientific Reports}\ }\textbf
  {\bibinfo {volume} {3}} (\bibinfo {year} {2013})}\BibitemShut {NoStop}%
\bibitem [{\citenamefont {Nuccio}\ \emph {et~al.}(2014)\citenamefont {Nuccio},
  \citenamefont {Schulz},\ and\ \citenamefont {Drew}}]{nuccio2014}%
  \BibitemOpen
  \bibfield  {author} {\bibinfo {author} {\bibfnamefont {L.}~\bibnamefont
  {Nuccio}}, \bibinfo {author} {\bibfnamefont {L.}~\bibnamefont {Schulz}}, \
  and\ \bibinfo {author} {\bibfnamefont {A.~J.}\ \bibnamefont {Drew}},\ }\href
  {\doibase 10.1088/0022-3727/47/47/473001} {\bibfield  {journal} {\bibinfo
  {journal} {Journal of Physics D: Applied Physics}\ }\textbf {\bibinfo
  {volume} {47}},\ \bibinfo {pages} {473001} (\bibinfo {year}
  {2014})}\BibitemShut {NoStop}%
\bibitem [{\citenamefont {Castelnovo}\ \emph {et~al.}(2010)\citenamefont
  {Castelnovo}, \citenamefont {Moessner},\ and\ \citenamefont
  {Sondhi}}]{castelnovo2010}%
  \BibitemOpen
  \bibfield  {author} {\bibinfo {author} {\bibfnamefont {C.}~\bibnamefont
  {Castelnovo}}, \bibinfo {author} {\bibfnamefont {R.}~\bibnamefont
  {Moessner}}, \ and\ \bibinfo {author} {\bibfnamefont {S.~L.}\ \bibnamefont
  {Sondhi}},\ }\href {\doibase 10.1103/PhysRevLett.104.107201} {\bibfield
  {journal} {\bibinfo  {journal} {Physical Review Letters}\ }\textbf {\bibinfo
  {volume} {104}} (\bibinfo {year} {2010}),\
  10.1103/PhysRevLett.104.107201}\BibitemShut {NoStop}%
\bibitem [{\citenamefont {Paulsen}\ \emph {et~al.}(2014)\citenamefont
  {Paulsen}, \citenamefont {Jackson}, \citenamefont {Lhotel}, \citenamefont
  {Canals}, \citenamefont {Prabhakaran}, \citenamefont {Matsuhira},
  \citenamefont {Giblin},\ and\ \citenamefont {Bramwell}}]{paulsen2014}%
  \BibitemOpen
  \bibfield  {author} {\bibinfo {author} {\bibfnamefont {C.}~\bibnamefont
  {Paulsen}}, \bibinfo {author} {\bibfnamefont {M.~J.}\ \bibnamefont
  {Jackson}}, \bibinfo {author} {\bibfnamefont {E.}~\bibnamefont {Lhotel}},
  \bibinfo {author} {\bibfnamefont {B.}~\bibnamefont {Canals}}, \bibinfo
  {author} {\bibfnamefont {D.}~\bibnamefont {Prabhakaran}}, \bibinfo {author}
  {\bibfnamefont {K.}~\bibnamefont {Matsuhira}}, \bibinfo {author}
  {\bibfnamefont {S.~R.}\ \bibnamefont {Giblin}}, \ and\ \bibinfo {author}
  {\bibfnamefont {S.~T.}\ \bibnamefont {Bramwell}},\ }\href {\doibase
  10.1038/nphys2847} {\bibfield  {journal} {\bibinfo  {journal} {Nature
  Physics}\ }\textbf {\bibinfo {volume} {10}},\ \bibinfo {pages} {135}
  (\bibinfo {year} {2014})}\BibitemShut {NoStop}%
\bibitem [{\citenamefont {Bovo}\ \emph {et~al.}(2013)\citenamefont {Bovo},
  \citenamefont {Bloxsom}, \citenamefont {Prabhakaran}, \citenamefont
  {Aeppli},\ and\ \citenamefont {Bramwell}}]{bovo2013}%
  \BibitemOpen
  \bibfield  {author} {\bibinfo {author} {\bibfnamefont {L.}~\bibnamefont
  {Bovo}}, \bibinfo {author} {\bibfnamefont {J.}~\bibnamefont {Bloxsom}},
  \bibinfo {author} {\bibfnamefont {D.}~\bibnamefont {Prabhakaran}}, \bibinfo
  {author} {\bibfnamefont {G.}~\bibnamefont {Aeppli}}, \ and\ \bibinfo {author}
  {\bibfnamefont {S.}~\bibnamefont {Bramwell}},\ }\href {\doibase
  10.1038/ncomms2551} {\bibfield  {journal} {\bibinfo  {journal} {Nature
  Communications}\ }\textbf {\bibinfo {volume} {4}},\ \bibinfo {pages} {1535}
  (\bibinfo {year} {2013})}\BibitemShut {NoStop}%
\bibitem [{\citenamefont {Liu}(1965)}]{liu1965}%
  \BibitemOpen
  \bibfield  {author} {\bibinfo {author} {\bibfnamefont {C.-T.}\ \bibnamefont
  {Liu}},\ }\emph {\bibinfo {title} {The effect of screening of the ionic
  atmosphere on the theory of the Wien effect in weak electrolytes}},\
  \href@noop {} {\bibinfo {type} {{Ph.D.}}},\ \bibinfo  {school} {Yale}
  (\bibinfo {year} {1965})\BibitemShut {NoStop}%
\end{thebibliography}%

\newpage\clearpage
\renewcommand\thefigure{S\arabic{figure}}
\setcounter{figure}{0}     
\section{Supplemental material}
\subsection{Response at higher temperatures}

\begin{figure}[htb!]
\centering
\includegraphics{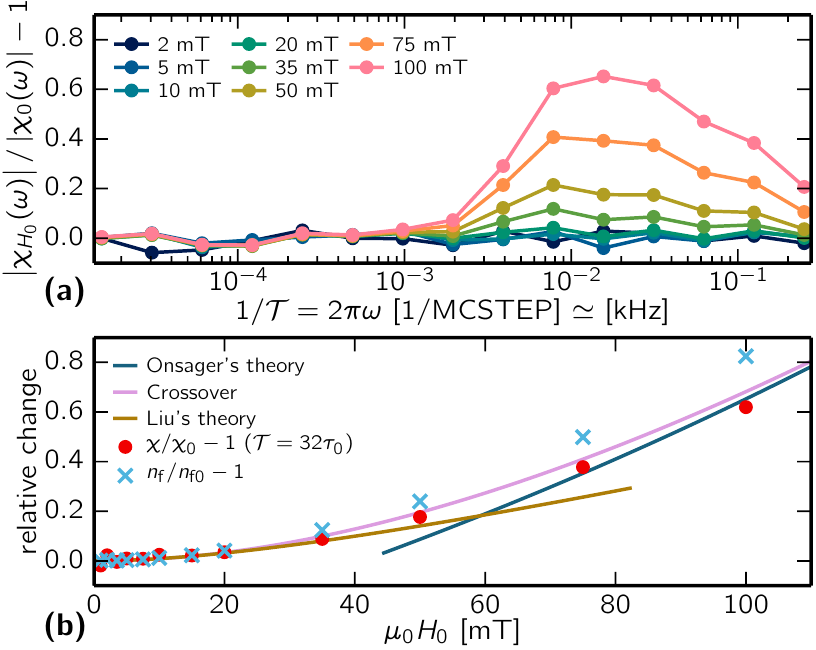}
\caption{\label{fig:hight} Non-linear susceptibility for DTO at $T=0.7~\mathrm{K}$ and demagnetization factor $\mathcal{D}=1/3$ ($n_\mathrm{tot}(0)\simeq 5.7\times10^{-3}$, and $n_\mathrm{f}(0)\simeq 4.4\times10^{-3}$). The high-temperature susceptibility requires a more precise choice of frequency and a detailed treatment of screening. (a) The non-linear susceptibility is visibly influenced by the Wien effect albeit at a narrower window of frequencies. (b) The amplitude dependence follows the prediction for the screened Wien effect given in Refs.~\cite{kaiser2013} (``Crossover'') and \cite{liu1965} (``Liu's theory''). }
\end{figure}

Thermalizing spin ice in the low temperature region, below $T_\mathrm{f}\simeq0.6\mathrm{K}$, is a difficult task. Moreover there are signs that different degrees of freedom thermalize at different time scales~\cite{snyder2001,snyder2004,slobinsky2010,erfanifam2011,pomaranski2013}, and it has not yet been established whether the hopping rate of $1\;\mathrm{kHz}$ extracted for spin ice above $T_\mathrm{f}$ \cite{jaubert2009} itself acquires
a temperature dependence below $T_\mathrm{f}$. If spin ice can be successfully thermalized at $0.45$ K one can expect the dumbbell model to provide an accurate description, with experiment yielding the non-linear magnetic response presented above. However, as shown in Fig.~\ref{fig:hight}, the Wien effect is present and measurable well above this temperature. In this regime screening has to be taken into account in detail (see below) as, using the parameters for DTO, the crossover field, which was less than 3 mT at 0.45 K, increases to 25 mT at 0.7 K (see Figs.~2,3,S1,S3). Results in Fig.~\ref{fig:hight} also include effects of demagnetization by taking it into account in the Ewald summation~\cite{melko2004}). Both demagnetization and increase in temperature act to narrow the frequency window for the Wien effect; yet it stays observable at $T=0.7~\mathrm{K}$ and demagnetization factor $\mathcal{D}=1/3$.

We note further that efficient heat extraction will be necessary during the field protocol  in order to avoid competing non-linear response induced by heating~\cite{castelnovo2010,paulsen2014}. Our model also neglects any minor adiabatic susceptibility response~\cite{bovo2013}.

\subsection{Screening: non-equilibrium activity coefficient}

Onsager's exact solution \cite{onsager1934} of the problem of two-charge dynamics yields an increase in dissociation constant $K(b)/K(0)=\sqrt{F(b)}$. This approach neglects many-body correlations, considering the free charge concentration to be that of a non-interacting gas with chemical potential $\nu$. This is valid for an electrolyte or magnetolyte as the charge concentration goes to zero, but away from this limit the long range interaction is screened, leading to an excess concentration. The common approach is to treat screening correlations in a mean-field fashion through the activity coefficient $\gamma$. In this case $K=\gamma^2n_\mathrm{f}^2/(2n_\mathrm{b})$ and the free charge density increase is taken to be
\begin{equation}
n_\mathrm{f}(b)/n_\mathrm{f}(0) = (\gamma(0)/\gamma(b)) \sqrt{F(b)} \,.
\end{equation}
The equilibrium coefficient, $\gamma(0)=\gamma_0$ can be calculated from electrolyte theory \cite{debye1923,bjerrum1926,moore1999}. At high field the screening charge cloud is swept away, as internal correlations among the free charges fail to establish if the ion drift velocity is too high. This is the first Wien effect which  allowed Onsager to set $\gamma(b)=1$ for fields above a threshold level, obtained by comparing the Debye screening length $\ell_\mathrm{D}=\sqrt{k_B T/\mu_0 Q_\mathrm{m}^2 \rho_\mathrm{f}}$ with $\ell_H=k_BT/\mu_0Q_\mathrm{m}H$, the length above which the field modifies the internal distribution of ions (see Fig.~S2). At low temperature the experimental range of fields extends well outside this threshold so that the effect is limited to a small offset to the unscreened theory \cite{kaiser2013}. At higher temperatures the threshold field is much larger and the crossover has to be considered in detail.

\begin{figure}[htb!]
\centering
\includegraphics[width=5.5cm]{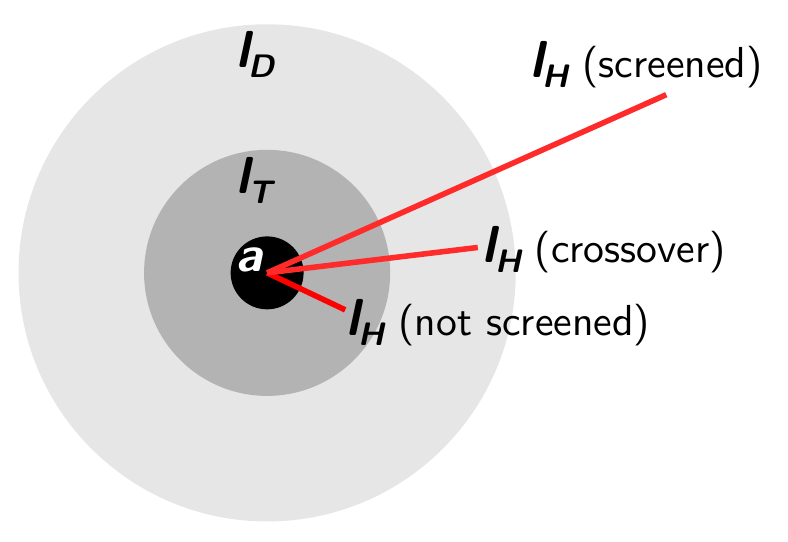}
\caption{\label{sfig:screening} At low fields ($\ell_H\gg\ell_\mathrm{D}$) the charges effectively interact with a screened potential with Liu's theory describing the increase in charge density. At high fields ($\ell_H\ll\ell_\mathrm{D}$) screening is destroyed and Onsager's theory holds. Crossover occurs in the field regime where $\ell_H\sim\ell_\mathrm{D}$. As neither of the two theories describes the crossover in full, we use a phenomenological treatment in terms of the non-equilibrium activity coefficient $\gamma(b)$. Also shown is the Bjerrum pairing length $\ell_T=\mu_0Q_\mathrm{m}^2/8\pi k_B T$.}
\end{figure}

We recently proposed a phenomenological non-equilibrium form for $\gamma(b)$, giving the charge increase at intermediate fields ($\ell_\mathrm{D}\sim\ell_H$) \cite{kaiser2013}.
We assumed  that the screening re-establishes a standard quadratic field dependence to leading order:~$n_\mathrm{f}(b)/n_\mathrm{f}(0) = 1+\mathrm{O} \left(b^2\right)$ and that the crossover to the high field limit, $\gamma(b\rightarrow\infty)=1$ occurs exponentially rapidly above the field threshold:
\begin{equation}
\gamma(b) = \left[1 + (1/\gamma_0-1)\exp{\left(-\frac{b}{2(1-\gamma_0)}\right)}\right]^{-1}\,.
\end{equation}
This formula accurately describes the simulation data in both magnetolytes and electrolytes and also compares favorably with Liu's \cite{liu1965}  screening corrections at low field. He argued that if $\ell_\mathrm{D}\ll\ell_H$, the effective two body potential is of the Debye form $U(r)\propto\exp(-r/\ell_\mathrm{D})/r$ which admits a perturbative solution for small field, yielding a non-equilibrium activity coefficient which can be cast in the form $\gamma_\mathrm{L}(b) = \gamma_0^{f_\mathrm{L}(\ell_\mathrm{D} / \ell_H)}$ with $f_\mathrm{L}(x) = \ln(1 + x)/x$. Data from our electrolyte simulations are compared with both expressions in Fig.~S3.

\begin{figure}[htb!]
\centering
\includegraphics[width=0.9\columnwidth]{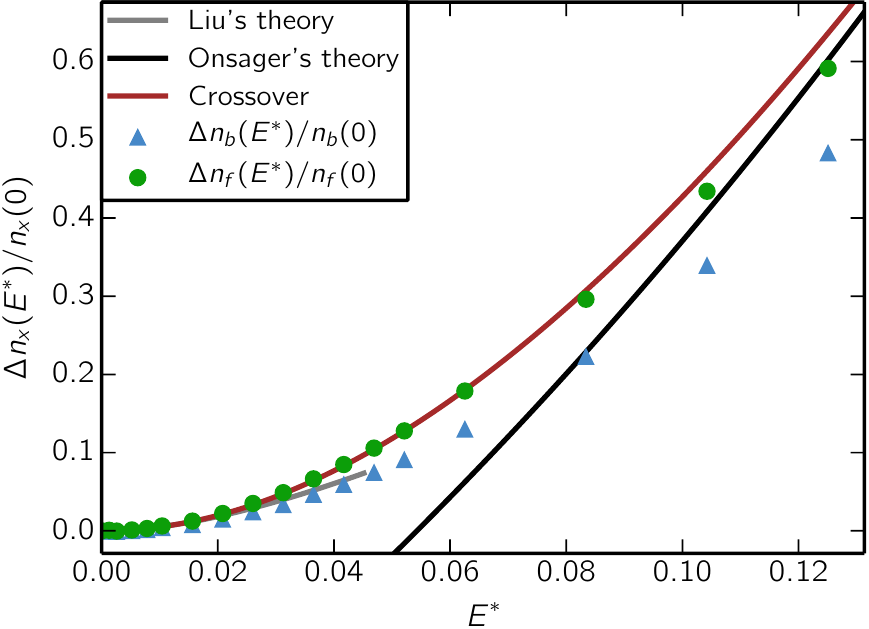}
\caption{\label{sfig:crossover} Effect of screening on the second Wien effect at higher densities are shown for a lattice Coulomb gas simulation at $\nu=-1.45\times U_\mathrm{C}(a)$ and $k_BT=0.212 \times U_\mathrm{C}(a)$, where $U_\mathrm{C}(a)$ is the nearest-neighbor Coulomb coupling; zero field densities are $n_\mathrm{f}(0)=1.24\times10^{-3}$ and $n_\mathrm{b}(0)=1.25\times10^{-3}$. Compare with a similar result for spin ice in Fig.~5. Liu's theory of diffusion in the Debye potential (grey line) describes the low field behavior and Onsager's unscreened expression (black line) is approached in high fields (diffusion in the Coulomb potential). The crossover is well characterised by a phenomenological theory (red line) that extends the activity coefficient out of equilibrium to describe the decay of the ionic atmosphere in an applied field. The free and bound charge populations are less clearly distinguished at higher densities and start to increase in a  similar manner.}
\end{figure}

\subsection{Quadratic reduction of density in field}

From Fig.~1c one can see that the long time, equilibrium monopole density for finite field, $n_\mathrm{f}^0(H_0)$, falls below that of zero field. The shift $n_\mathrm{f}^0(H_0) - n_\mathrm{f}^0(0)$ varies quadratically with applied field. This effect appears already for a single tetrahedron as the energy costs for monopole creation by flipping a spin with, or against, the field differ by the Zeeman energy. By symmetry, the terms linear in field cancel, so that 
\begin{align}
n_\mathrm{f}^0(H_0) \simeq n_\mathrm{f}^0 \left(1-\frac{2}{3}\left(\frac{\mu_0 \mathbf{\mu}\cdot \mathbf{H}_0}{k_B T}\right)^2\right) \,.
\end{align}
The Wien effect, \emph{linear} in $H_0$, occurs as, out of equilibrium, the magnetic field provides a chemical potential gradient for the monopoles, in analogy with the electrolyte. In this environment the symmetry principle is violated, only to be re-established at long time as the system returns to equilibrium. 

The initial conditions for the kinetic model therefore depend on the selected experimental protocol. A sample prepared in zero field would correspond to zero initial magnetization $m(0)=0$ and zero excess charge $\zeta(0)=0$. 
However, if measured with respect to equilibrium in finite field, $\zeta(t=0)$ has to be modified accordingly. It is necessary to replace the initial zero-field condition $\zeta(0)=0$ with
\begin{align} \label{eq:ch5:q_init}
\zeta(0) = \zeta_0 = \frac{n_\mathrm{f}(0)-n_\mathrm{f}^0(H_0)}{n_\mathrm{f}(H_0)-n_\mathrm{f}^0(H_0)} \,.
\end{align}
where $n_\mathrm{f}(H_0)$ denotes the steady-state Wien effect free charge density. We also reinterpret $\zeta(t)$ as the relative increase in density from the in-field equilibrium base line towards the steady state value. Note that as the equilibrium charge decrease is quadratic in field, it  does not require a direct modification of the kinetic model which contains only terms linear in field. 

\begin{figure}[htbp!]
\centering
\includegraphics[width=0.9\columnwidth]{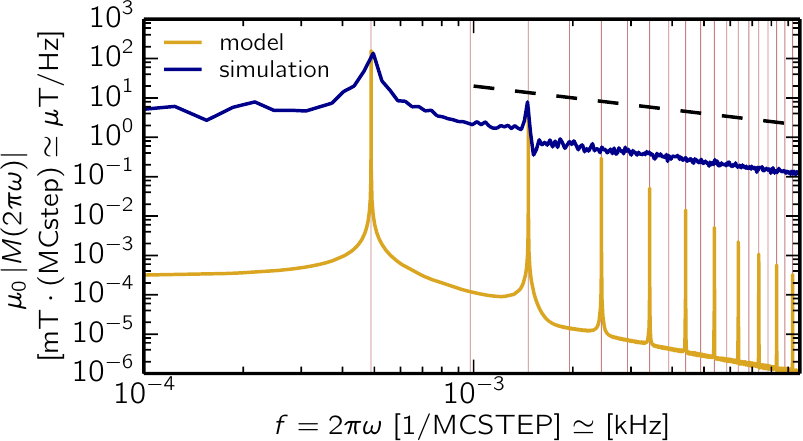}
\caption{\label{fig:spectrum} Spectral density of the magnetization. Parameters are for DTO at $0.45~\mathrm{K}$, driven by a sinusoidal field with $\mathcal{T}=2048\tau_0$ and $\mu_0H_0=50\;\mathrm{mT}$. We observe response at the first and third harmonic on a background of thermal fluctuations. The background has a Debye-like form and is therefore consistent with the susceptibility via the fluctuation-dissipation theorem (dashed black line $\propto 1/\omega$). The kinetic model (yellow line) gives the correct peak weight; it is however athermal. Higher harmonics (at thin red lines) predicted by the kinetic model are below the thermal background in this simulation.}
\end{figure}

\subsection{Spectral analysis}

We first record the time trace of magnetization ($\sim10$ samples) for every amplitude and frequency in our simulations. Afterwards, we keep only the part corresponding to the periodic steady-state and discard the equilibration process. We perform a Fourier analysis to obtain the first and higher non-linear susceptibilities defined in the main text (see Fig.~\ref{fig:spectrum}). We use zero-padding and a Hamming window to improve the quality of the results, although the choice of window has little influence on the extracted values of the first non-linear susceptibility. This also allows us to observe higher harmonics and confirm that only odd harmonics are present in the non-linear response. The highest harmonic observed was the $9^\mathrm{th}$, when using parameters corresponding to  DTO at $T=0.45\;\mathrm{K}$, $\mu_0H_0=100\;\mathrm{mT}$, and $\mathcal{T}=2048\;\mathrm{MC\;steps}$ (not shown).

\end{document}